\documentclass[11pt]{article}%\usepackage{palatino,url,geometry}
\usepackage{url,geometry}
\usepackage{subcaption}
\usepackage{graphics,latexsym,amsfonts,verbatim,amsmath}
\usepackage{algorithm}
\usepackage{algpseudocode}
\usepackage{todonotes}
\usepackage{lineno}
\usepackage{multirow}
\usepackage{tikz}
\usepackage{pbox}
\usepackage{bibentry}
\usepackage{bm}
\usetikzlibrary{shapes,arrows}
\usepackage{yhmath}
        \usepackage{setspace}
\usepackage{authblk}

\usepackage{booktabs}
\usepackage{paralist}
\usepackage{colortbl}
\usepackage{xcolor}
\usepackage[bookmarksnumbered=true]{hyperref}
\usepackage{natbib}
\usepackage{amssymb}

\def\hat{\widehat}

\newcommand{\exclude}[1]{}
\newtheorem{lemma}{Lemma}

\newtheorem{assumption}{Assumption}
\newtheorem{proposition}{Proposition}

\newcommand{\nx}[1]{{\color{blue}{#1}}}

\tikzstyle{box} = [rectangle, rounded corners, minimum width=1cm, minimum height=0.5cm, text centered, text width=3cm, draw=black]
\tikzstyle{box2} = [rectangle, rounded corners, minimum width=1cm, minimum height=0.5cm, text centered, text width=2cm, draw=black]

\tikzstyle{arrow} = [thick,->,>=stealth]

\title{Collaborative Analysis for Paired A/B Testing Experiments}

\author[1]{Qiong Zhang}
\affil[1]{School of Mathematical and Statistical Sciences, Clemson University}

\author[2]{Lulu Kang}
\affil[2]{Department of Mathematics and Statistics,  University of Massachusetts Amherst}

\author[3]{Xinwei Deng}
\affil[3]{Department of Statistics, Virginia Tech}
\date{}

\begin{document}
\maketitle

\begin{abstract}
With the extensive use of digital devices, online experimental platforms are commonly used to conduct experiments to collect data for evaluating different variations of products, algorithms, and interface designs, a.k.a., A/B tests.
In practice, multiple A/B testing experiments are often carried out based on a common user population on the same platform. 
The same user's responses to different experiments can be correlated to some extent due to the individual effect of the user. 
In this paper, we propose a novel framework that collaboratively analyzes the data from paired A/B tests, namely, a pair of A/B testing experiments conducted on the same set of experimental subjects. 
The proposed analysis approach for paired A/B tests can lead to more accurate estimates than the traditional separate analysis of each experiment. 
We obtain the asymptotic distribution of the proposed estimators and demonstrate that the proposed estimators are asymptotically the best linear unbiased estimators under certain assumptions. 
%Theoretical justification is also established to demonstrate the merits of the proposed method. 
%\textcolor{red}{Maybe more concrete on what "theoretical justification", such as "asymptotic distribution of ... so that we obtain ...".}
Moreover, the proposed analysis approach is computationally efficient, easy to implement, and robust to different types of responses.
Both numerical simulations and numerical studies based on a real case are used to examine the performance of the proposed method. 
\end{abstract}


\section{Introduction}

\subsection{Background and Motivation}
With the global coverage of the internet, many online platforms, offering e-commerce, digital services, social media, etc., have been established by technology companies, governments, healthcare and education organizations. 
They have engaged a huge population of users. 
Massive amounts of data are recorded daily from user activities, serving as the basis of data-driven decisions and policies. 
This potential has prompted these organizations to collect data through online platforms more proactively by carrying out online controlled experiments (e.g., A/B testing experiments). 
As pointed out by \cite{kohavi2020trustworthy}, A/B testing experiments are often effective in revealing new insights to generate significant impacts potentially, and thus A/B testing has become ubiquitous in these organizations. 

\begin{figure}[ht]
\centering
\includegraphics[width=0.9\textwidth]{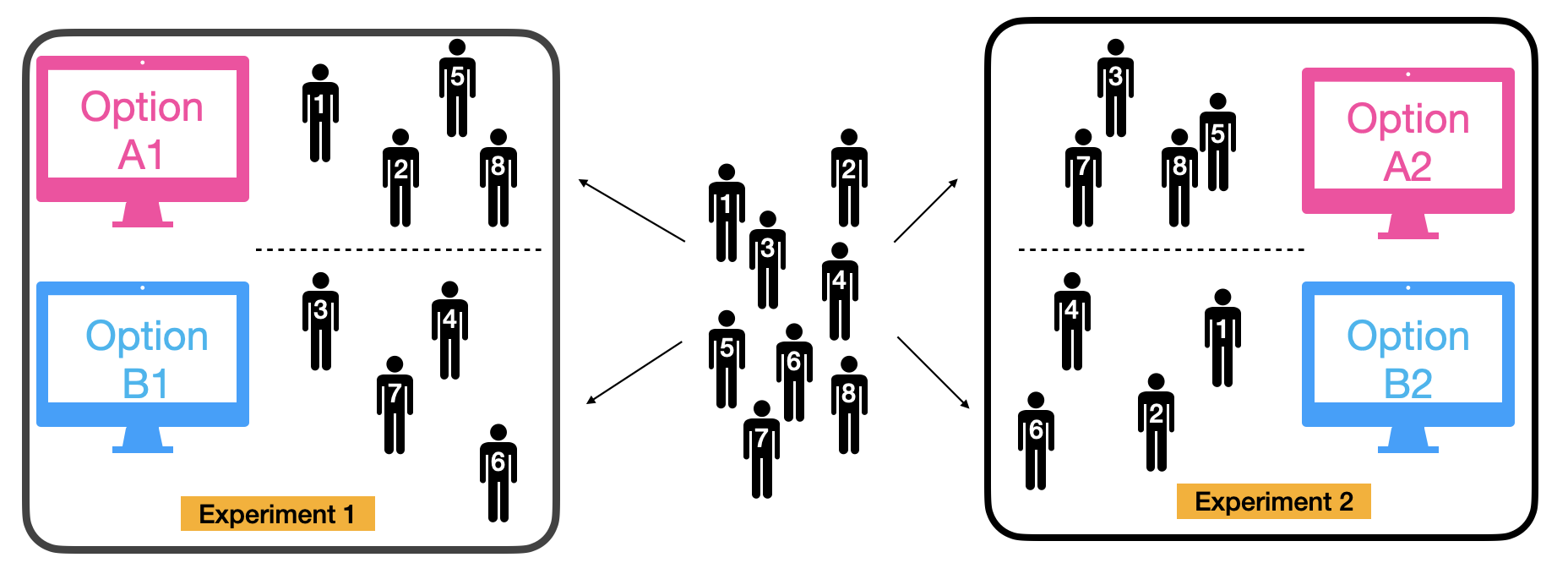}
\caption{An illustration of paired A/B tests: eight users are participating in two experiments, each with two levels. The users' responses to both experiments are recorded.}\label{fg:coe}
\end{figure}

For an online experimental platform, multiple A/B test experiments are often conducted on the same population of users within a short time frame, especially when the treatment settings involved in different experiments are not in conflict \citep{nassi_jewkes_2021}. 
In this paper, we consider the paired A/B tests, namely, 
a pair of A/B testing experiments conducted on the same set of experimental subjects during the same time frame. 
As illustrated in Figure \ref{fg:coe}, the same set of users are participating in both experiments.  
Each experiment compares two different treatment settings, i.e., A1/B1 or A2/B2.
In each experiment, users are divided into two groups based on the treatment settings that they are assigned to, and their responses to both experiments are recorded. 
Since the outcomes of the pair of experiments from the same user share common user characteristics, 
the analysis of experiments can be more effective if they are conducted collaboratively rather than separately for each experiment.

We illustrate the concept of paired (or multiple) A/B tests using the example from the online math teaching platform, ASSISTments \citep{heffernan2014assistments}, also see \url{https://new.assistments.org/}. 
This platform supports ``E-trials'' for different teaching methods. 
\cite{selent2016assistments} described a set of experiments based on this platform (early version). 
According to \cite{selent2016assistments}, 
a group of students participated in 
multiple ASSISTments experiments. 
Each experiment is associated with solving a type of math problem such as multiplying mixed numbers, equivalent expressions, and writing inequalities from situations, etc. 
Different types of problems can be created by different math teachers. 
Students in each experiment were randomly assigned to two hint conditions (treatment and control) to solve a sequence of math problems of the same type. 
The outcomes can be whether or not the student completed this type of problem (binary), the number of problems the students solved before completing (counts), the total time spent on this type of problem (continuous), etc. 
One research goal is to estimate the treatment effects for each type of math problem and find out the hint condition that leads to the best student performances for each particular problem type. 
One student is often involved in more than one experiment (e.q., type of math problems). 
In this case, the students' characteristics make an important contribution to their experimental outcomes. 
Therefore, combining experiments that share common users can potentially enhance the analysis of the treatment effect of one individual experiment. 
This example also indicates that different from one $2\times 2$ factorial experiment, paired A/B testing experiments are two experiments with different sets of control and treatment, possibly conducted by different project teams on the same online experimental platform.

\subsection{Related Literature}

Thanks to its wide application in practice, there has been a growing interest in the statistical research community to study the new challenges of A/B testing and online controlled experiments. 
The combined literature on this topic from statistics, machine learning, and application areas has attracted great attention.
The review paper by \cite{larsen2023statistical} provides a comprehensive look into the new challenges and development of online controlled experiments and A/B testing in the recent literature. 
Due to the space limit, we only highlight a few works closely related to our work. 
The potential outcome framework \citep{neyman1923application, rubin1974estimating, Imbens_Rubin_2015, ding2024first} is the foundation for conducting causal inference of A/B testing experiments.
To accommodate users' heterogeneity, covariate measurements of users can be available from prior experiments. 
\cite{deng2013improving}, \cite{poyarkov2016boosted} and \cite{jin2023toward} provided linear and nonlinear ways to adjust the outcome measurements based on covariates to reduce the variance of the estimate. 
When covariate measurements are hidden, \cite{syrgkanis2019machine} and \cite{zhao2022adjust} proposed different ways to estimate the treatment effect. 
It is also noted that recent works considered to make inferences on the treatment effect with missing outcomes \citep{chen2015semiparametric, shen_mao_zhang_chen_nie_deng_2023, zuo2024mediation, zhao2024covariate}. 
For multiple A/B experiments,  \cite{gupta2019top} described the common scenario that users are simultaneously involved in hundreds of A/B testing experiments, which is the challenge this work aims to overcome.
Despite the large literature on A/B testing experiments, few works consider how to jointly analyze multiple A/B testing experiments. 

\subsection{Our Contribution and Paper Outline}

In this paper, we aim to fill the gap of jointly analyzing paired (or multiple) A/B testing experiments to discover new knowledge that was buried when each experiment was analyzed separately.
To start, we focus on the paired A/B testing experiments.
Particularly, we consider each user has unique characteristics that are not necessarily available as covariate measurements to the experimenter. 
The proposed collaborative estimators of the treatment effects are easy to compute and can be quickly implemented in 
 any large-scale online platform. Our method can be extended to the cases with 
missing outcomes for a common practice: partially paired experiments.   
Theoretically and numerically, we demonstrate that the proposed estimators that combine the information of the paired experiments are better than the traditionally used separate experiment analysis approach. 
We also perform a robustness analysis to show the numerical performances of the proposed method when the model assumption is misspecified. The proposed collaborative analysis framework
 improves the design and analysis of future A/B testing practices and makes better use of all the simultaneous experiments. 

The paper is organized as follows. Section \ref{sec:exact} illustrates the proposed method under the ideal situation: complete paired experiments with balanced and orthogonal designs. Section \ref{sec:partial} extends the method to partially paired experiments with nearly balanced and orthogonal designs. Section \ref{sec:num} provides numerical studies and Section \ref{sec:case} provides a case study with a real problem background. Section \ref{sec:con} concludes the paper with a discussion and future directions. 

%\section{Fully Paired A/B Tests with Orthogonal Designs}\label{sec:exact}

\section{Collaborative Analysis of Fully Paired A/B Tests}\label{sec:exact}

Assume that there are two experiments each with a two-level experimental factor, i.e., paired A/B tests, and they are conducted on the same group of experimental units (i.e., experimental subject). 
The goal is to estimate the treatment effect for each of the two experimental factors. 
Let $y_{i,k}$ be the experimental outcome of the $i$-th unit of the $k$-th experiment with $k=1, 2$ and $i=1,\ldots, n$.
The underlying model of the outcome is assumed to be  
\begin{equation}\label{eq:true_model}
y_{i,k}=u_i+\alpha_k+x_{i,k} \beta_k+\epsilon_{i,k}, \quad k=1,2,
\end{equation}
where $\alpha_k$ is the intercept, $x_{i,k}\in\{-1, 1\}$ is the design setting for $i$-th unit and $\beta_k$ is the treatment effect of the $k$-th factor, and $\epsilon_{i,k}$ is the random noise that is independent of test units, with mean zero and variance $\sigma^2_k$.
The individual effect $u_i$ can be a random effect with mean zero and variance $\tau^2$, independent of the random noise.
It represents the unique characteristics of each unit and carries the dependence of the outcomes from a paired of experiments.
We do not assume any probabilistic distributions for $u_i$'s and $\epsilon_{i,k}$'s. 
Note that, combining the models of the two experiments in \eqref{eq:true_model} forms the linear mixed effect model (see for example, \cite{andrzej2012linear}) with individual effect $u_i$. 
In this paper, we formulate the problem of paired A/B testings under the framework of linear mixed effect models. 

\subsection{Analysis of Paired A/B Tests}
Under the model assumption in \eqref{eq:true_model}, we illustrate the ideas of a collaborative analysis of paired A/B tests by describing three analysis approaches to estimate the treatment effects. 
For clarity of illustration, we consider the ideal case that 
the \emph{balanced} and \emph{orthogonal} designs are used for the two experiments, as stated in Assumption \ref{ass:orth}.
\begin{assumption}\label{ass:orth}
Assume that the balanced and orthogonal designs are used for the two experiments, i.e.,
\begin{equation}\label{eq:orth}
\sum^n_{i=1}x_{i,1}=\sum^n_{i=1}x_{i,2}=\sum^n_{i=1}x_{i,1}x_{i,2}=0.
\end{equation}
and the level combinations of the two design factors $x_{i,1}$ and $x_{i,2}$ form a random partition of the $n$ users into four groups. 
\end{assumption}
\emph{Remark 1.} We are aware that the orthogonal design assumption rarely holds in practice, but it facilitates our initial introduction of the proposed estimator. 
Later we will discuss the situation in Section \ref{sec:partial} where this assumption is relaxed. 

\paragraph{Single Analysis:}
In this paper, single analysis refers to the case that each of the two experiments is analyzed separately. 
Under Assumption \ref{ass:orth}, the least squared estimator, denoted by $\hat{\beta}_k^s$, for each single experiment is given by 
\begin{equation}\label{eq:single}
\hat\beta_k^s=n^{-1}\sum^{n}_{i=1}x_{i,k}y_{i,k}\quad\mathrm{and}\quad
\mathrm{Var}(\hat\beta^s_k)=\frac{\tau^2+\sigma^2_k}{n},\quad\mathrm{for}\quad k=1,2.
\end{equation}
%Similarly, for the second experiment, we have that
%\[
%\hat\beta_2^s=n^{-1}\sum^{n}_{i=1}x_{i2}y_{i2}\quad\mathrm{and}\quad
%\mathrm{Var}%(\hat\beta^s_2)=\frac{\tau^2+\sigma^2_2}{n}.
%\]
Also, notice that $\mathrm{Cov}\left(\hat\beta_1^s, \hat\beta_2^s\right)=\tau^2 n^{-2}\sum^n_{i=1}x_{i,1}x_{i,2}=0$.

%If we further consider the normality assumptions of the random effects $u_i$, $\epsilon_{i,1}$ and $\epsilon_{i,2}$, the estimators $\hat\beta^s_1$ and $\hat\beta^s_2$ are independent and normally distributed, which can simplify the inference of any contrast (linear combination of $\beta_1$ and $\beta_2$) based on the two estimators. 

\paragraph{Paired Analysis:}
For the paired experiments with common user random effects in \eqref{eq:true_model}, one can bypass the random effects $u_i$ by taking the differences of the two outcomes from the same user:
\begin{equation}\label{eq:linear_diff}
    z_i \triangleq y_{i,1}-y_{i,2}=\alpha+x_{i,1}\beta_1-x_{i,2}\beta_2+\delta_i,
\end{equation}
where $\alpha=\alpha_1-\alpha_2$ and $\delta_i=\varepsilon_1-\varepsilon_2$. 
Under this model \eqref{eq:linear_diff}, the estimators of the treatment effects $\beta_1$ and $\beta_2$ can be obtained by the least squared regression method. Let $\bm{x}_1=(x_{1,1}, \ldots, x_{n,1})^\top$ and $\bm{x}_2=(x_{1,2}, \ldots, x_{n,2})^\top$ be the design vectors of the paired experiments. The design matrix of this model is
$X=(\bm{1}_n, \bm{x}_1, -\bm{x}_2)$. Then we have that the least squared estimator of $\bm \theta=(\alpha, \beta_1, \beta_2)^\top$ is
\begin{equation}\label{eq:lse}
\hat {\bm \theta}=(\hat\alpha, \hat\beta^p_1, \hat\beta^p_2)^\top=(X^\top X)^{-1} X^\top\bm z,
\end{equation}
where $\bm z=(z_1, \ldots, z_n)^\top$. 
The analysis based on the model of the differences of the two outcomes is called \emph{paired analysis}.
%Based on the differences of responses, the difference-in-mean estimator of $\beta_1$ is given by
%\begin{equation}\label{eq:dm_paired}
%    \tilde\beta^d_1=\frac{1}{2}\left(\frac{\sum^n_{i=1}z_{i}I(x_{i1}=1)}{\sum^n_{i=1}I(x_{i1}=1)}-\frac{\sum^n_{i=1}z_{i}I(x_{i1}=-1)}{\sum^n_{i=1}I(x_{i1}=-1)}\right),
%\end{equation}
%with
%\[
%\mathrm{Var}(\hat\beta^d_1)=\frac{\sigma^2_1+\sigma^2_2}{n\left(1-\frac{\bar x^2_1+\bar x^2_2-2\bar x_1\bar x_2n^{-1}\sum^n_{i=1}x_{i1}x_{i2}}{1-\bar x^2_2}\right)},
%\]
%and
%\[
%\mathrm{Var}(\tilde\beta^d_1)=\frac{1}{4}\left(\frac{1}{\sum^n_{i=1}I(x_{i1}=1)}+\frac{1}{\sum^n_{i=1}I(x_{i1}=-1)}\right)(\sigma^2_1+\sigma^2_2).
%\]

Under Assumption \ref{ass:orth}, the least square solution leads to paired analysis estimators, denoted by $\hat{\beta}_k^p$,
\begin{equation}\label{eq:diff}
\hat\beta^p_1=n^{-1}\sum^n_{i=1}x_{i,1}z_i
\quad\mathrm{and}\quad \hat\beta^p_2=-n^{-1}\sum^n_{i=1}x_{i,2}z_i
\end{equation}
and
\[
\mathrm{Var}\left(\hat\beta^p_1\right)=\mathrm{Var}\left(\hat\beta^p_2\right)=\frac{\sigma^2_1+\sigma^2_2}{n}\quad\mathrm{and}\quad\mathrm{Cov}\left(\hat\beta^p_1, \hat\beta^p_2\right)=n^{-2}(\sigma^2_1+\sigma^2_2)
\sum^n_{i=1}x_{i,1}x_{i,2}=0.
\]
The relative efficiency between the single analysis estimator and paired analysis estimator is
\begin{equation}\label{eq:RE}
\mathrm{RE}\left(\hat\beta^s_k, \hat\beta^p_k\right)=\frac{\sigma^2_1+\sigma^2_2}{\sigma^2_k+\tau^2}\quad\mathrm{for}\quad
k=1,2.
\end{equation}
We see that the estimators based on the paired experiments are more efficient than the estimator of the single experiment analysis if $\tau^2>\sigma^2_k$. %Otherwise, the estimator based on the single experiment is more efficient. 
Since both estimators are unbiased estimators of $\beta_k$ under the model assumption in \eqref{eq:lse}, we can combine them to obtain another unbiased estimator of the treatment effect which is shown to be universally more efficient than both the single and paired analysis, as introduced next. 

\paragraph{Proposed Collaborative Analysis:} Given the two unbiased estimators $\hat\beta^s_k$ and $\hat\beta^p_k$ of $\beta_k$, we can obtain another unbiased estimator of $\beta_k$ by taking a linear combination of $\hat\beta^s_k$ and $\hat\beta^p_k$. 
The linear weights are given by Lemma \ref{lemma1}. 
In this paper, we call this analysis approach \emph{collaborative analysis} and the resulting estimator \emph{collaborative estimator}. 

\begin{lemma}\label{lemma1}
Suppose that $\bm T$ is a $d\times 1$ random vector with mean $\mu \bm 1_d$ and variance-covariance matrix $\bm \Sigma$. Then the best unbiased linear estimator of $\mu$ given by $\bm T$ is
\begin{equation}\label{eq:blue}
\hat\mu=\frac{\bm 1^\top_d\bm \Sigma^{-1}\bm T}{\bm 1^\top_d\bm \Sigma^{-1}\bm 1_d}
\end{equation}
where the linear weights are given by $\bm \Sigma^{-1}\bm 1_d/\bm 1^\top_d\bm \Sigma^{-1}\bm 1_d$, which is derived from solving the optimization problem:
\[\mathrm{min}_{\bm a\in \mathbb{R}^d} \bm a^\top \bm \Sigma\bm a
\quad\mathrm{s.t.}\quad \bm a^\top \bm 1_d=1.
\]
\end{lemma}

Consider $\hat\beta^s_k$ and $\hat\beta^p_k$ as a $d=2$ random vector whose
mean vector and covariance matrix of $(\hat\beta^s_k, \hat\beta^p_k)^\top$ are 
\[
\beta_k\bm 1_2
\quad \mathrm{and}\quad
\frac{1}{n}\left[\begin{array}{cc}
\tau^2+\sigma^2_k & \sigma^2_k\\
\sigma^2_k & \sigma^2_1+\sigma^2_2
\end{array}
\right]\quad\mathrm{for}\quad k=1,2,
\]
respectively. 
Lemma \ref{lemma1} gives the best linear weights to construct 
an unbiased estimator of $\beta_k$ based on the single and paired analysis estimators. The resulting collaborative estimators, denoted by $\hat{\beta}^c_k$, are given by
\begin{equation}\label{eq:collaborative}
\hat\beta^c_1=\frac{\tau^2 \hat\beta^p_1+\sigma^2_2 \hat\beta^s_1}{\tau^2+\sigma^2_2}\quad\mathrm{and}
\quad \hat\beta^c_2=\frac{\tau^2 \hat\beta^p_2+\sigma^2_1 \hat\beta^s_2}{\tau^2+\sigma^2_1}
\end{equation}
with $\mathrm{cov}(\hat\beta^c_1, \hat\beta^c_2)=0$, and variances
\[
\mathrm{Var}\left(\hat\beta^c_1\right)=\frac{\sigma^2_1\tau^2+\sigma^2_2\tau^2+\sigma^2_1\sigma^2_2}{n(\sigma^2_2+\tau^2)}\quad\mathrm{and}
\quad \mathrm{Var}\left(\hat\beta^c_2\right)=\frac{\sigma^2_1\tau^2+\sigma^2_2\tau^2+\sigma^2_1\sigma^2_2}{n(\sigma^2_1+\tau^2)}.
\]
Thus, the relative efficiencies over the single and paired analysis estimators are
\[
\mathrm{RE}\left(\hat\beta^s_k, \hat\beta^c_k\right)=\frac{\sigma^2_1\tau^2+\sigma^2_2\tau^2+\sigma^2_1\sigma^2_2}{(\sigma^2_1+\tau^2)(\sigma^2_2+\tau^2)}\leq 1,\quad\mathrm{RE}\left(\hat\beta^p_k, \hat\beta^c_k\right)=\frac{(\sigma^2_k+\tau^2)}{(\sigma^2_1+\sigma^2_2)}\cdot\frac{\sigma^2_1\tau^2+\sigma^2_2\tau^2+\sigma^2_1\sigma^2_2}{(\sigma^2_1+\tau^2)(\sigma^2_2+\tau^2)}\leq 1,
\]
which indicates that the collaborative estimator is more efficient than the single and paired analysis estimators disregarding the values of the parameters $\tau^2$, $\sigma^2_1$, and $\sigma^2_2$. 

Note that the proposed collaborative estimators contain unknown parameters $\tau^2$, $\sigma^2_1$, and $\sigma^2_2$. 
We propose the following plug-in estimators for them. 
Let $S^2_{k+}$ and $S^2_{k-}$ be the sample variances of the $k$-th experiment under experimental setting 1 and -1, respectively. 
Let $S^2_{++}$, $S^2_{+-}$, $S^2_{-+}$ and $S^2_{--}$ be the sample variances of $z_i$'s in \eqref{eq:linear_diff} under experimental setting $x_{i,1}$ and $x_{i,2}$ with the sub-index representing the signs of $x_{i,1}$ and $x_{i,2}$. 
We obtain the moment estimators 
\begin{equation}\label{eq:v12}
\widehat{\sigma^2_k+\tau^2}=\frac{S^2_{k+}+S^2_{k-}}{2}\quad\mathrm{for}\quad k=1,2,
\end{equation}
and
\begin{equation}\label{eq:v3}
\widehat{\sigma^2_1+\sigma^2_2}=\frac{S^2_{++}+S^2_{+-}+S^2_{-+}+S^2_{--}}{4}.
\end{equation}
Therefore, the moment estimators of $\tau^2$, $\sigma^2_1$ and $\sigma^2_2$ can be calculated by solving the linear systems \eqref{eq:v12}-\eqref{eq:v3}:
\begin{equation}\label{eq:moment}
\left[\begin{array}{c}
  \hat\tau^2\\
   \hat\sigma^2_1 \\
   \hat\sigma^2_2
\end{array}\right]=
\frac{1}{2}
\left[\begin{array}{rrr}
1 & 1& -1\\
1 & -1& 1\\
-1 & 1& 1
\end{array}\right]
\left[\begin{array}{c}
 \widehat{\sigma^2_1+\tau^2}\\
   \widehat{\sigma^2_2+\tau^2} \\
  \widehat{\sigma^2_1+\sigma^2_2}
\end{array}\right]
\end{equation}

In sum, to construct the collaborative estimators under Assumption \ref{ass:orth}, we will need to first compute $\hat\beta^s_k$ and $\hat\beta^p_k$,  $S^2_{k+}$ and $S^2_{k-}$ for $k=1,2$, and $S^2_{++}$, $S^2_{+-}$, $S^2_{-+}$ and $S^2_{--}$, and then assemble them according to \eqref{eq:collaborative} and \eqref{eq:moment}. Under the model assumption, we can also obtain maximum likelihood estimators of the variance parameters. The computation of the moment estimators is trivial and straightforward to implement. 

\subsection{Properties of Collaborative Estimators for Paired A/B Tests}

Although the proposed collaborative estimators are the best linear combination of the single and paired analysis estimators, we have not fully demonstrated the benefit of the proposed approach compared to other potential estimators. 
The following proposition further states that the collaborative estimators are the best unbiased linear estimators of $\beta_1$ and $\beta_2$ with respect to the original responses vector $(y_{i,1}, y_{i,2})$'s in \eqref{eq:true_model}. 
\begin{proposition}\label{prop:exact}
Assume that $\tau^2$, $\sigma^2_1$ and $\sigma^2_2$ are known. Under the model assumption in \eqref{eq:true_model} and the balance and orthogonal assumption of the two designs in \eqref{eq:orth}, the collaborative estimators $\hat\beta^c_1$ and $\hat\beta^c_2$ in \eqref{eq:collaborative} have the following properties:
\begin{itemize}
\item[(i)] They are the best unbiased linear estimators (BLUE) of $\beta_1$ and $\beta_2$.

\item[(ii)]  $\hat\beta^c_1$ and $\hat\beta^c_2$ are uncorrelated, and 
\[
\left(\frac{\hat\beta^c_1-\beta_1}{\sqrt{\mathrm{Var}(\hat\beta^c_1)}}, \frac{\hat\beta^c_2-\beta_2}{\sqrt{\mathrm{Var}(\hat\beta^c_2)}}\right)^\top\rightarrow \mathcal N_2(\bm 0_2, \bm I_2)
\]
in distribution, as $n\rightarrow \infty$.
\end{itemize}
\end{proposition}
The proof of this proposition is deferred to Appendix \ref{app:prop1}.
The proof also demonstrates that under Assumption \ref{ass:orth}, the proposed collaborative estimators echo the weighted least squared estimators, which are known as the BLUE
under model \eqref{eq:true_model}. 
The proposed collaborative estimators are convenient to compute and attain clear asymptotic properties. 
In Sections \ref{sec:partial}, we further relax the assumptions in Section \ref{sec:exact} and extend this approach to partially paired A/B tests. 

\section{Collaborative Analysis of Partially Paired A/B Tests}\label{sec:partial}

Often, the experimental outcomes of some experimental units may not be available. 
Since the missing outcomes are not planned in the experimental design stage, the orthogonality assumption \ref{ass:orth} can not hold. 
In this section, we extend the collaborative analysis approach to partially paired A/B tests.

%If a unit has no outcome observed, this unit should be removed from the analysis. So we assume that each unit has at least one outcome observed. 
Without loss of generality, we assume that,  for $i=1, \ldots, n_0$, the outcomes of both experiments from the $i$-th unit are available;
for unit $i=n_0+1, \ldots, n_0+n_1$, the outcome of the first experiment is available, and for unit $i=n_0+n_1+1, \ldots, n_0+n_1+n_2$, the outcome of the second experiment is available; 
for unit $i=n_0+n_1+n_2+1,\ldots, n$, the outcomes of both experiments are missing.
The structure of collected data is illustrated in Table \ref{tb1}.
\begin{table}[!h]
\begin{center}
\caption{The structure of data collected from paired experiments}\label{tb1}
\begin{tabular}{|c|cc|cc| }
\hline
  & \multicolumn{2}{c|}{Experiment 1} & \multicolumn{2}{c|}{Experiment 2} \\\cline{2-5}
  Unit & Outcome & Design &  Outcome & Design \\\hline
 $1$ & $y_{1,1}$ & $x_{1,1}$  & $y_{1,2}$ & $x_{1,2}$ \\
  $2$  & $y_{2,1}$& $x_{2,1}$   & $y_{2,2}$ & $x_{2,2}$ \\
  $\vdots$ & $\vdots$ & $\vdots$ & $\vdots$ & $\vdots$\\
  $n_0$  & $y_{n_0,1}$& $x_{n_0,1}$  & $y_{n_0,2}$ & $x_{n_0,2}$ \\\hline
 $n_0+1$ & $y_{n_0+1,1}$ & $x_{n_0+1,1}$ & NA & $x_{n_0+1,2}$ \\
  $n_0+2$ & $y_{n_0+2,1}$ & $x_{n_0+2,1}$ & NA & $x_{n_0+2,2}$ \\
  $\vdots$ & $\vdots$ & $\vdots$ & $\vdots$ & $\vdots$\\
  $n_0+n_1$ & $y_{n_0+n_1,1}$ &$x_{n_0+n_1,1}$ & NA & $x_{n_0+n_1,2}$ \\\hline
 $n_0+n_1+1$ &NA& $x_{n_0+n_1+1,1}$  & $y_{n_0+n_1+1,2}$ & $x_{n_0+n_1+1,2}$ \\
  $n_0+n_1+2$ &NA& $x_{n_0+n_1+2,1}$  & $y_{n_0+n_1+2,2}$ & $x_{n_0+n_1+2,2}$ \\
  $\vdots$ &NA& $\vdots$  & $\vdots$ & $\vdots$\\
  $n_0+n_1+n_2$ &NA& $x_{n_0+n_1+n_2,1}$  & $y_{n_0+n_1+n_2,2}$ & $x_{n_0+n_1+n_2,2}$ \\\hline
   $n_0+n_1+n_2+1$ &NA& $x_{n_0+n_1+n_2+1,1}$  & NA & $x_{n_0+n_1+n_2+1,2}$ \\
 % $n_0+n_1+2$ &NA& $y_{n_0+n_1+2,2}$  & $x_{n_0+n_1+2,1}$ & $x_{n_0+n_1+2,2}$ \\
  $\vdots$ &$\vdots$& $\vdots$  & $\vdots$ & $\vdots$\\
  $n$ &NA&$x_{n,2}$ & NA & $x_{n,2}$ 
  \\\hline
\end{tabular}
\end{center}
\vspace{-20pt}
\end{table}

Given the structure of collected data in Table \ref{tb1}, we insert the following assumptions on the collected data.
\begin{assumption}\label{ass:n}
As $n\rightarrow\infty$, \[
\frac{n_0}{n}\rightarrow r_0\in (0, 1],\quad
\frac{n_1}{n}\rightarrow r_1\in [0, 1),\quad
\frac{n_2}{n}\rightarrow r_2\in [0, 1),
\]
with $r_0$, $r_1$ and $r_2$ being constants satisfying that $r_0+r_1\in (0, 1]$ and $r_0+r_2\in (0, 1]$. 
 \end{assumption}
Assumption \ref{ass:n} ensures that there is a sufficient number of observations for each experiment. Therefore, the corresponding treatment effects are estimable. 
%If the total number of observations from one of the experiments is negligible compared to the total number of users, the data collection has some issues, and the analysis of that experiment can be misleading.
Assumption \ref{ass:n} also guarantees that the number of units in the fully paired portion of the experiment is sufficiently large to ensure the necessity of collaborative estimation.
A few special cases are: 1) if $r_0=1$, the two experiments are exactly overlapped, which is the same as the situation described in Section \ref{sec:exact}; 2) if $r_1=0$ or $r_2=0$, one experiment is ``nested'' in the other experiment, but the following proposed collaborative analysis still works.

\begin{assumption}\label{ass:nearly}
The designs of the paired experiments are nearly 
 balanced and orthogonal, namely, 
\[
\sum^{n_0}_{i=1}x_{i,1}=\sum^{n_0}_{i=1}x_{i,2}=
\sum^{n_0+n_1}_{i=n_0+1}x_{i,1}=
\sum^{n_0+n_1+n_2}_{i=n_0+n_1+1}x_{i,2}=
\sum^{n_0}_{i=1}x_{i,1}x_{i,2}=o(n).
\]
\end{assumption}
Assumption \ref{ass:nearly} can be easily satisfied if the original design of experiments can be well controlled under Assumption \ref{ass:orth} or each assignment of $x_{i,k}$ is completely randomized and independent of each other for $k=1,2$ plus that the responses are missing at random and independent with the design allocation. 

We now construct the collaborative estimators based on partially paired data in Table \ref{tb1}.
Based on the fully paired outcomes for units 1 to $n_0$ from the top panel in Table \ref{tb1}, we can construct the paired estimator of $\beta_1$ as in \eqref{eq:diff}:
\[
n^{-1}_0\sum^{n_0}_{i=1}x_{i,1}z_i\quad\mathrm{with}\quad z_i=y_{i,1}-y_{i,2}.
\]
Also, we can obtain two single analysis estimators from the top panel and the second panel respectively. In sum, 
 we have that the following three estimators for the estimand $\beta_1$:
\begin{equation}\label{eq:three}
n^{-1}_0\sum^{n_0}_{i=1}x_{i,1}z_i, \quad n^{-1}_0\sum^{n_0}_{i=1}x_{i,1}y_{i,1}, \quad \mathrm{and}\quad n^{-1}_1\sum^{n_0+n_1}_{i=n_0+1}x_{i,1}y_{i,1}.
\end{equation}
Under the model assumption in \eqref{eq:true_model} and Assumptions \ref{ass:n} and \ref{ass:nearly},
all three estimators are asymptotic unbiased estimators of $\beta_1$. 
Also, the covariance matrix of three estimators in \eqref{eq:three} is proportional to
\[
\left[
\begin{array}{ccc}
n^{-1}_0(\sigma^2_1+\sigma^2_2) & n^{-1}_0\sigma^2_1 & 0\\
n^{-1}_0\sigma^2_1 & n^{-1}_0(\sigma^2_1+\tau^2) & 0\\
0 & 0 & n^{-1}_1(\sigma^2_1+\tau^2)
\end{array}
\right].
\]
The covariance matrix also indicates that the weights of the two single analysis estimators are different. So we should separate the two single analysis estimators as two estimators in collaborative analysis.
Based on Lemma \ref{lemma1}, we can combine three estimators to obtain the \emph{proposed collaborative estimator} of $\beta_1$ for the partially paired case:
\begin{equation}\label{eq:coe}
\hat{\beta}^c_1=\frac{ \left.\sum\limits^{n_0}_{i=1}x_{i,1}z_i\middle/\left(\sigma^2_1+\sigma^2_2+\frac{\sigma^2_1\sigma^2_2}{\tau^2}\right)\right.+ \left.\sum\limits^{n_0}_{i=1}x_{i,1}y_{i,1}\middle/\left(\tau^2+\sigma^2_1+\frac{\sigma^2_1\tau^2}{\sigma^2_2}\right)\right.+ \left.\sum\limits^{n_0+n_1}_{i=n_0+1}x_{i,1}y_{i,1}\middle/\left(\tau^2+\sigma^2_1\right)\right.}{\left(\sigma^2_1+\sigma^2_2+\frac{\sigma^2_1\sigma^2_2}{\tau^2}\right)^{-1}n_0+ \left(\tau^2+\sigma^2_1+\frac{\sigma^2_1\tau^2}{\sigma^2_2}\right)^{-1}n_0+(\tau^2+\sigma^2_1)^{-1}n_1},
\end{equation}
where the weights are derived according to Lemma \ref{lemma1} to ensure that $\hat{\beta}^c_1$ is the best linear combination of the three estimators in \eqref{eq:three}
under the model assumption in \eqref{eq:true_model}.  
We also have that
\begin{equation}\label{eq:var_coe}
\mathrm{Var}\left( 
\hat{\beta}^c_1\right)=\left[\left(\sigma^2_1+\sigma^2_2+\frac{\sigma^2_1\sigma^2_2}{\tau^2}\right)^{-1}n_0+ \left(\tau^2+\sigma^2_1+\frac{\sigma^2_1\tau^2}{\sigma^2_2}\right)^{-1}n_0+(\tau^2+\sigma^2_1)^{-1}n_1\right]^{-1}.
\end{equation}
For comparison purposes, we provide the results for the single analysis estimator and the paired analysis estimators under partially paired experiments as follows:
\[
\hat\beta^s_1=\frac{
\sum^{n_0}_{i=1}x_{i,1}y_{i,1}+\sum^{n_0+n_1}_{i=n_0+1}x_{i,1}y_{i,1}}{n_0+n_1}\quad \mathrm{and}\quad\mathrm{Var}\left(\hat\beta^s_1\right)=\frac{\tau^2+\sigma^2_1}{n_0+n_1},
\]
\[
\hat\beta^p_1=n^{-1}_0
\sum^{n_0}_{i=1}x_{i,1}z_{i}\quad \mathrm{and}\quad\mathrm{Var}\left(\hat\beta^p_1\right)=\frac{\sigma^2_1+\sigma^2_2}{n_0}.
\]
We see that if $\tau^2\rightarrow\infty$, the estimator $\hat{\beta}^c_1$ becomes the paired analysis estimator $\hat{\beta}^p_1$, and if $\tau^2\rightarrow 0$, this estimator becomes the single analysis estimator $\hat{\beta}^s_1$.
Similar to $\hat{\beta}^c_1$, the collaborative estimator of $\beta_2$ is
\begin{equation}\label{eq:coe2}
\hat{\beta}^c_2=\frac{ \left.\sum\limits^{n_0}_{i=1}x_{i,2}z_i\middle/\left(\sigma^2_1+\sigma^2_2+\frac{\sigma^2_1\sigma^2_2}{\tau^2}\right)\right.+ \left.\sum\limits^{n_0}_{i=1}x_{i,2}y_{i,2}\middle/\left(\tau^2+\sigma^2_2+\frac{\sigma^2_2\tau^2}{\sigma^2_1}\right)\right.+ \left.\sum\limits^{n_0+n_1+n_2}_{i=n_0+n_1+1}x_{i,2}y_{i,2}\middle/\left(\tau^2+\sigma^2_2\right)\right.}{\left(\sigma^2_1+\sigma^2_2+\frac{\sigma^2_1\sigma^2_2}{\tau^2}\right)^{-1}n_0+ \left(\tau^2+\sigma^2_2+\frac{\sigma^2_2\tau^2}{\sigma^2_1}\right)^{-1}n_0+(\tau^2+\sigma^2_2)^{-1}n_2},
\end{equation}
with
\[
\mathrm{Var}(\hat{\beta}^c_2)=\left[\left(\sigma^2_1+\sigma^2_2+\frac{\sigma^2_1\sigma^2_2}{\tau^2}\right)^{-1}n_0+ \left(\tau^2+\sigma^2_2+\frac{\sigma^2_2\tau^2}{\sigma^2_1}\right)^{-1}n_0+(\tau^2+\sigma^2_2)^{-1}n_2\right]^{-1}.
\]

Different from the fully paired A/B tests under Assumption \ref{ass:orth} in Section \ref{sec:exact}, the collaborative estimators for partially paired A/B tests are no longer the BLUEs under model \eqref{eq:true_model}. 
However, the collaborative estimators are still the asymptotically best linear unbiased estimators under model \eqref{eq:true_model} as shown in the following proposition. 
\begin{proposition}\label{prop:partial}
Suppose that $\tau^2$, $\sigma^2_1$ and $\sigma^2_2$ are known. Under the model assumption in \eqref{eq:true_model}, data structure in Table \ref{tb1} and Assumptions \ref{ass:n}-\ref{ass:nearly}, the collaborative estimators in \eqref{eq:coe} and \eqref{eq:coe2} have the following properties:
\begin{itemize}
\item[(i)] They are asymptotically the best linear unbiased estimators of the treatment effects $\beta_1$ and $\beta_2$  under the true model assumption \eqref{eq:true_model};
%i.e., it is asymptotically equivalent to the weighted least squared estimators
\item[(ii)] The asymptotic distribution of $\hat\beta^c_1$
and $\hat\beta^c_2$ in Proposition \ref{prop:exact} also holds under the partially paired case.  
\end{itemize}
\end{proposition}
The proof of this proposition is deferred to Appendix \ref{app:prop2}.

The collaborative estimators for partially paired A/B tests also contain unknown parameters $\tau^2$, $\sigma^2_1$ and $\sigma^2_2$, which can be estimated as in \eqref{eq:v12}, \eqref{eq:v3} and \eqref{eq:moment}. However, different from the fully paired case in Section \ref{sec:exact}, the sample variances 
$S^2_{k+}$ and $S^2_{k-}$ for $k=1,2$, and $S^2_{++}$,  $S^2_{+-}$, $S^2_{-+}$ and $S^2_{--}$ are computed only based on available outcomes. 
We summarize the steps of the collaborative analysis procedure for partially paired experiments as follows.
\begin{itemize}
\item[\bf Step 1.] For $k=1, 2$, obtain sample variances $S^2_{k+}$ and $S^2_{k+}$ using available outcomes $y_{i,k}$ from experiment $k$ with design value equal to 1 and -1, respectively.   
\item [\bf Step 2.] Split the differences $z_i=y_{i,1}-y_{i,2}$'s from the paired parts, i.e., $i=1, \ldots, n_0$ in Table \ref{tb1} according to the designs $x_{i,1}$ and $x_{i,2}$ 
into four groups and obtain sample variances $S^2_{++}$, $S^2_{+-}$, $S^2_{-+}$ and $S^2_{--}$
of each group, respectively.
\item [\bf Step 3.] Compute the moment estimators: 
$\widehat{\sigma^2_1+\tau^2}$, $\widehat{\sigma^2_2+\tau^2}$, 
and $\widehat{\sigma^2_1+\sigma^2_2}$ according to 
\eqref{eq:v12} and \eqref{eq:v3}.
\item [\bf Step 4.] Compute the variance estimators 
$\hat \sigma^2_1$, $\hat\sigma^2_2$ and $\hat\tau^2$
according to \eqref{eq:moment}.
\item [\bf Step 5.] Obtain the collaborative estimators $\hat\beta^c_k$ for $k=1,2$ according to \eqref{eq:coe} and \eqref{eq:coe2} with the variance estimators from Step 4. 
\end{itemize}
The computational complexity of the whole procedure is $\mathcal O(n)$, which is the same as the estimators of the single and paired analysis.

\section{Numerical Study}\label{sec:num}

We provide numerical studies to compare the proposed method with other alternatives. All the methods involved in the comparison are described below: 
\begin{itemize}
\item[1.] \textbf{SINGLE:} The single experiment estimator in \eqref{eq:single} with available outcomes from a single experiment.
\item[2.] \textbf{PAIRED:} The least squared estimator based on the difference model in \eqref{eq:diff} with fully paired outcomes.
%\item[2.] \textbf{WLS:} The weighted least squared estimator in Appendix \ref{app:prop2} with plug-in weight estimates.
\item[3.]\textbf{COE:} The proposed collaborative estimator in \eqref{eq:coe} with all available outcomes.
\item[4.]\textbf{LME:} The estimators are given by fitting a linear mixed-effects model using the \texttt{R} function ``lmer'' in the package \texttt{lme4} \citep{lme4}.
The model is assumed to be \eqref{eq:true_model} with the normal random effects $u_i$'s and normal random errors $\varepsilon_{i,k}$'s. Using this function, the unknown variance parameters are given by optimizing the restricted maximum likelihood (REML) criterion in this package \citep{lme4}. %\textcolor{red}{This acronym needs to be explained and citations should be added.}
\end{itemize}
It is worth noting that, the estimators generated by LME are equivalent to the weighted least squared estimators with variance parameters $\tau^2$, $\sigma^2_1$ and $\sigma^2_2$ given by maximum likelihood type approaches. Therefore, using the same set of plug-in variance estimators, the difference of the estimators given by LME and COE converges to zero as $n\rightarrow\infty$
given by the proof of Proposition \ref{prop:partial} in the Appendix. 
%\textcolor{red}{I am not entirely sure, but we might need to add a little more detailed explanation (or derivations) to back up this claim. It might not be transparent to readers.}

We describe the data generation scheme for the simulation study as follows.
Assume that there are $n$ test units. 
The outcomes of the paired experiments are simulated from the model 
\begin{equation}\label{eq:simulation_model}
%\tilde y_{ik}(x)=g\left(y_{ik}(x)\right)\quad \mathrm{with}\quad
y_{i,k}(x)=1+x \beta_k+\epsilon_{i,k}+u_i,\quad\mathrm{for}\quad i=1, \ldots, n, \quad k=1,2, \quad\mathrm{and}\quad x\in\{-1, 1\},
\end{equation}
where $x \beta_k$ is a linear function of $x$, 
$\epsilon_{i,k}$ is a mean-zero normal random error with variance $\sigma^2_k=1$ and $u_i$ is the individual random effect. 
We use some different ways to generate $u_i$'s.
If $u_i$'s are iid samples from $N(0, \tau^2)$, the outcomes are exactly generated under the model assumption in \eqref{eq:true_model}. 
We generate two $n\times 1$ design vectors $\bm x_1$ and $\bm x_2$ satisfying the balanced and orthogonal assumption in Assumption \ref{ass:orth}. 
For $k=1,2$, we generate the responses $y_{i,k}(x)$ for each entry in $\bm x_k$.
We specify the missing rate of outcomes $r$ and randomly mark $n r$ responses as missing for each experiment. 

We use mean squared error (MSE) to evaluate the accuracy of the estimators. 
Without loss of generality, we only report the MSE of $\beta_1$ for the first experiment.
The results and conclusion for $\beta_2$ should be similar to $\beta_1$.
For an estimator, we replicate the data generation and estimation procedure under the same setting 100 times and obtain 100 estimators $\hat\beta^1_1, \ldots, \hat\beta^{100}_1$. 
%Let $\tilde\beta_1$ be the estimand of the first experiment. 
For a clear comparison of different estimators, we compute the ratio between the MSEs of one estimator and the single experiment estimator:
\begin{equation}\label{eq:evaluation}
\mathrm{MSE.ratio}=\frac{100^{-1}\sum^{100}_{l=1}(\hat\beta^l_1-\beta_1)^2}{100^{-1}\sum^{100}_{l=1}(\hat\beta^{s,l}_1-\beta_1)^2},
\end{equation}
where $\hat\beta^{s,l}_1$'s are the single experiment estimators of $\beta_1$ in \eqref{eq:single}. The value of $\mathrm{MSE.ratio}$ is the smaller the better. If this value is less than 1, it indicates that the estimator is more accurate than the single analysis estimator. 

In the remaining of this section, we conduct numerical studies to investigate three aspects: (1) the computational advantage of COE compared with LME; (2) the robustness of COE under different assumptions of the user effect $u_i$ in \eqref{eq:simulation_model}; (3) the robustness of COE under different types of responses. 
%by transferring the original response $y_{i,k}(x)$ in \eqref{eq:simulation_model}. %, i.e., different variations of the function $g(\cdot)$ in \eqref{eq:simulation_model}.

\subsection{Computational Advantage}

\begin{figure}[ht]
\centering
\includegraphics[width=0.8\textwidth]{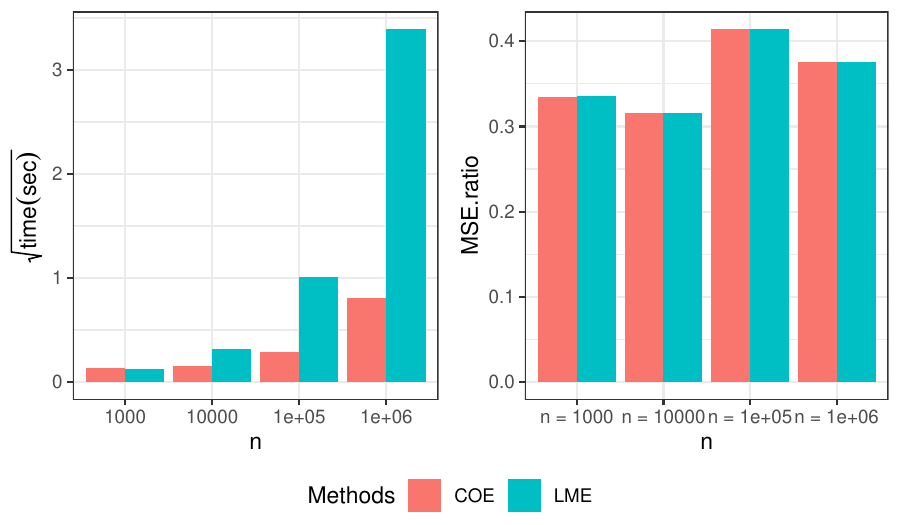}
\caption{Comparison between COE and LME on the average computational time over 100 replication and the MSE.ratio in \eqref{eq:evaluation}.}\label{fig:time}
\end{figure}

We demonstrate the computational advantage of COE compared with LME. 
The responses are generated following model in \eqref{eq:simulation_model} with  $u_i\overset{\text{iid}}{\sim}N(0, \tau^2)$ with $\tau=2$ and $\beta_k=1$ for $k=1, 2$. Therefore, the data is exactly generated by the true model in \eqref{eq:true_model}. 
Since the purpose of this study is to demonstrate the computational advantage, we assume that the data are exactly paired without missing values. 
To compute the MSE.ratio in \eqref{eq:evaluation}, $\beta_1=1$. 
We show the MSE.ratio and average computational time (i.e., average CPU time over 100 replications) in Figure \ref{fig:time}.
According to the results in Figure \ref{fig:time}, COE outperforms LME in computational time, and the
advantage of COE is more distinct as $n$ increases.
Also, COE and LME have the same level of accuracy, and both outperform the single experimental estimator since the values of MSE.ratio are less than one. 
%\textcolor{red}{Need to define "computational time". Is it "CPU time"?}

\subsection{Robustness to Random Effect}

We consider different settings of the random effect $u_i$'s. 
For $i=1, \ldots, n$, we generate $\bm w_i\overset{\text{iid}}{\sim}MVN(0, I_{10})$, which is fixed for each user and used as the user's ten-dimensional latent covariates. 
\begin{itemize}
\item[(a)] We generate $u_i\overset{\text{iid}}{\sim}N(0, \tau^2)$
under the true model assumption in \eqref{eq:true_model}. 
Under this setting, the random effect is not associated with the latent covariates $\bm w_i$. We expect that COE has the best performance as stated in the theoretical results. Also, as demonstrated in \eqref{eq:RE}, PAIRED 
outperforms SINGLE when the value of $\tau$ (e.q., $\tau/\sigma_2$ since $\sigma_2=1$ in the simulation) is greater than one, whereas SINGLE outperforms PAIRED when the value of $\tau$ is less than one. 

\item[(b)] We generate 
\[
u_i=\bm w^\top_i\bm \gamma\quad\mathrm{with}\quad
\bm\gamma\sim MVN(0, \tau^2 I_{10}).
\]
Under this setting, the variances (conditional on $\bm w_i$) of the random effects are different for different test units. Although the data under this case is not generated as the true model in \eqref{eq:true_model},  we expect that taking the difference between the outcomes of two experiments can still remove the user effects. Therefore, both COE and PAIRED can show some advantages especially when the value of $\tau$ is larger than 1. 

%\textcolor{red}{$\var(u_i)=\var(\bm w_i^\top \bm \gamma)=\E[\var(\bm w_i^\top \bm \gamma|\bm \gamma)]+\var[\E(\bm w_i^\top \bm \gamma|\bm \gamma)]=\E(\bm \gamma^\top \bm \gamma)+0$. Same for different test units.}
\item[(c)] We generate
\[
u_i=\bm w^\top_i\bm \gamma_k\quad\mathrm{with}\quad
\bm\gamma_k\sim MVN(0, \tau^2 I_{10})
\quad\mathrm{for}\quad k=1, 2.
\]
Under this setting, the user effects of the same individual are different for different experiments. Therefore, taking the difference between the outcomes of the two experiments can not remove the user effects. We expect that PAIRED can not outperform SINGLE. However, the performances of COE remain robust and should at least be similar to SINGLE.  
%\item[d.] We generate 
%\[
%u_i=(\bm w^\top_i\bm \gamma_k)^2\quad\mathrm{with}\quad
%\gamma_k\sim MVN(0, \tau^2 I_{10})
%\quad\mathrm{for}\quad k=1, 2.
%\]
%Under this setting, the variances of the user effects are different for different users. The user effects of the same individual are different for different experiments. Also, the user effect is non-linear with respect to the latent covariates $\bm w_i$. 
\item[(d)] We first generate $\bm\gamma_k\sim MVN(0, \tau^2 I_{10})$ for $k=1,2$, then generate
\[
u_i=\bm w^\top_i\bm \gamma_1 I(x_{i,k}=1)+\bm w^\top_i\bm \gamma_2I(x_{i,k}=-1)
\quad\mathrm{for}\quad k=1, 2.
\]
where $x_{ik}$ is the treatment allocation of $i$-th user in the $k$-th experiments. Under this setting, there are interactions between the user and treatment effects. Taking the difference between the outcomes of the two experiments may have a small contribution in removing the user effects due to the interaction. Therefore, we expect that PAIRED can not outperform SINGLE, but COE can outperform SINGLE slightly.
\end{itemize}

\begin{figure}[t!]
\centering
        \includegraphics[scale=0.8]{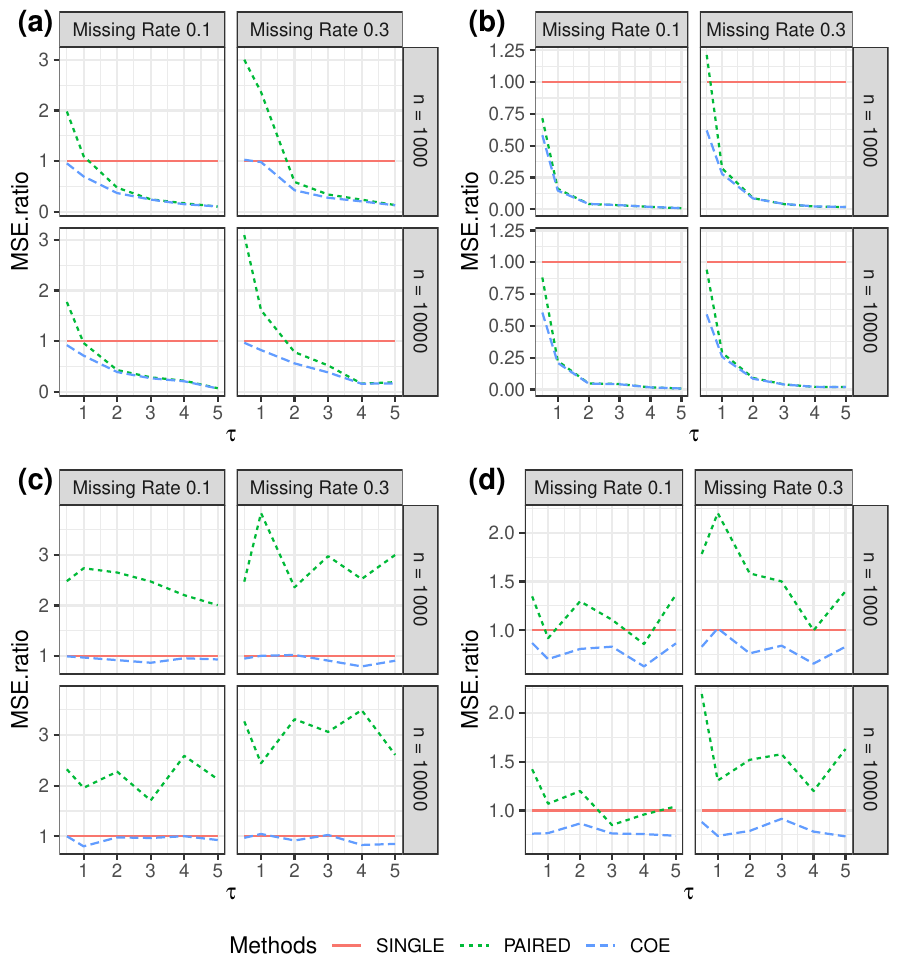}
\caption{MSE.ratio in \eqref{eq:evaluation} under user effect settings (a)-(d)}\label{fig:user}
\end{figure}

For all four settings, we specify $\beta_k=1$ for $k=1,2$. We vary the standard deviation of the user effect $\tau\in \{0.5, 1, 2, 3, 4, 5\}$, the missing rate $r\in \{0.1, 0.3\}$ and sample size $n\in \{1000, 10000\}$. 
The results of $\text{MSE.ratio}$ are shown in Figure \ref{fig:user}. Similarly to what we expected when introducing each case above, the results show that the MSE.ratio of COE is consistently the best among the three methods for all four random effects settings $u_i$. Compared to SINGLE, the advantage of COE is larger for larger variance parameter $\tau^2$. Compared to PAIRED, the advantage of COE is larger for smaller variance parameter $\tau^2$ or larger missing rate. Also, we can see that COE is more robust than PAIRED with respect to misspecified models (i.e., case (c) and (d)).
Also, the advantage of COE in terms of MSE.ratio does not diminish as we increase the sample size $n$.

\subsection{Robustness to Different Types of Outcomes}

Although the derivation of the COE is for continuous outcome data, we show here that the COE outperforms other estimators for discrete outcome data as well. 
In this part, we consider binary and integer types of outcomes. 
For convenience, we generate the discrete outcomes based on some latent continuous response that follows the model in \eqref{eq:true_model}. 
Specifically, for a binary outcome, we define it based on the continuous outcome $y_{i,k}(x)$ from \eqref{eq:true_model} as 
\begin{equation}\label{eq:binary}
\tilde y_{i,k}(x)=I\left\{y_{i,k}(x)>\text{median}\right\},
\end{equation}
where ``median'' is the median outcomes of all $y_{i1}(1)$ for $i=1, \ldots, n$. 
For an integer or count outcome, we define it based the continuous outcome $y_{i,k}(x)$ from \eqref{eq:true_model} as 
\begin{equation}\label{eq:integer}
\tilde y_{i,k}(x)=\left \lfloor{(y_{i,k}(x)-\text{min})^{1/2}}\right \rfloor,
\end{equation}
where $\left \lfloor{t}\right \rfloor$ gives the largest integer that is less than or equal to $t$ and ``min'' is the minimum of $y_{i,k}$ for $i=1, \ldots, n$, $k=1, 2$ and $x\in \{-1,1\}$. 

For the discrete outcomes, the estimand in such cases is the average treatment effect, defined as $\tilde{\beta}_1$ below, which is usually the default parameter of interest in causal inference \citep{Imbens_Rubin_2015}
\[
\tilde\beta_1=\frac{\sum^n_{i=1}\left\{\tilde y_{i,1}(1)-\tilde y_{i,1}(-1)\right\}}{2n},
\]
which is computed for each replication. Then the MSE.ratio in \eqref{eq:evaluation} is modified by
\begin{equation}\label{eq:evaluationModified}
\mathrm{MSE.ratio}=\frac{100^{-1}\sum^{100}_{l=1}(\hat\beta^l_1-\tilde\beta^l_1)^2}{100^{-1}\sum^{100}_{l=1}(\hat\beta^{s,l}_1-\tilde\beta^l_1)^2}.
\end{equation}
Under the model for continuous outcome in \eqref{eq:true_model}, the expected value of $\tilde\beta^l_1$ coincides with $\beta_1$. 
Therefore, our previous theoretical and numerical results still hold when the average treatment effect is the parameter of interest, provided the underlying model is \eqref{eq:true_model}.

\begin{figure}[t!]
\centering

        \includegraphics[scale=0.8]{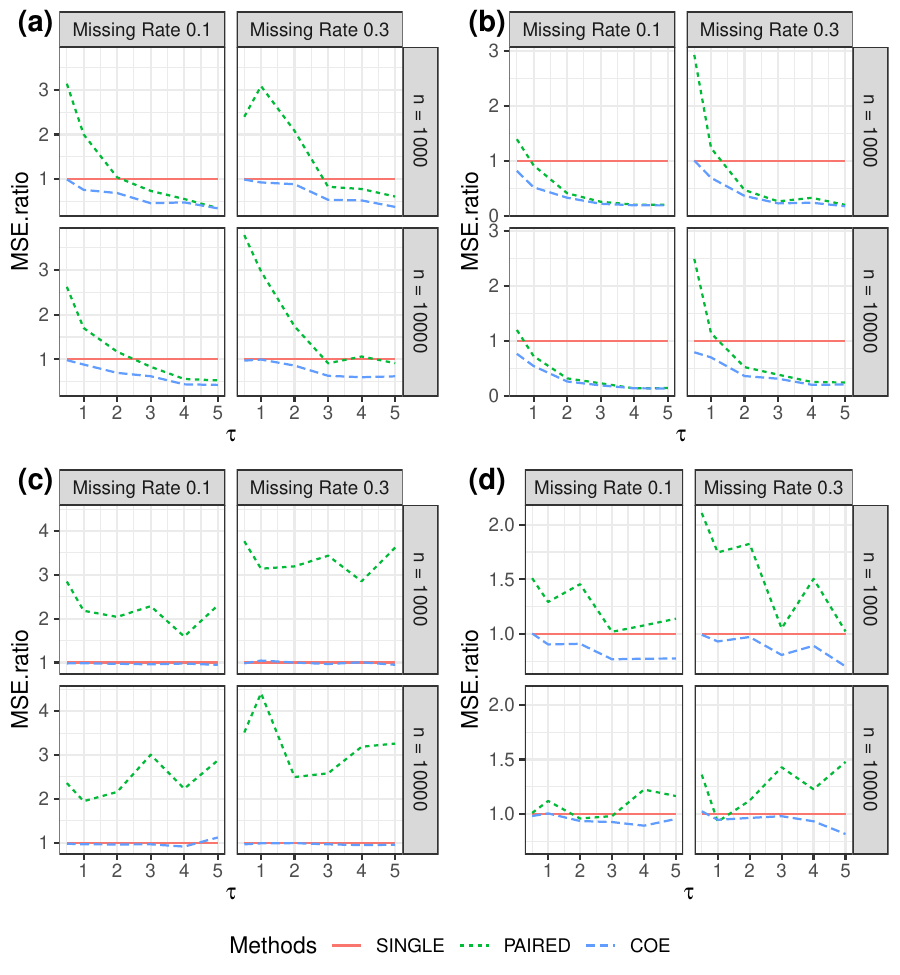}

\caption{Modified MSE.ratio in \eqref{eq:evaluationModified} under user effect settings (a)-(d) for binary responses.}\label{fig:binary}
\end{figure}

\begin{figure}[t!]
\centering
        \includegraphics[scale=0.8]{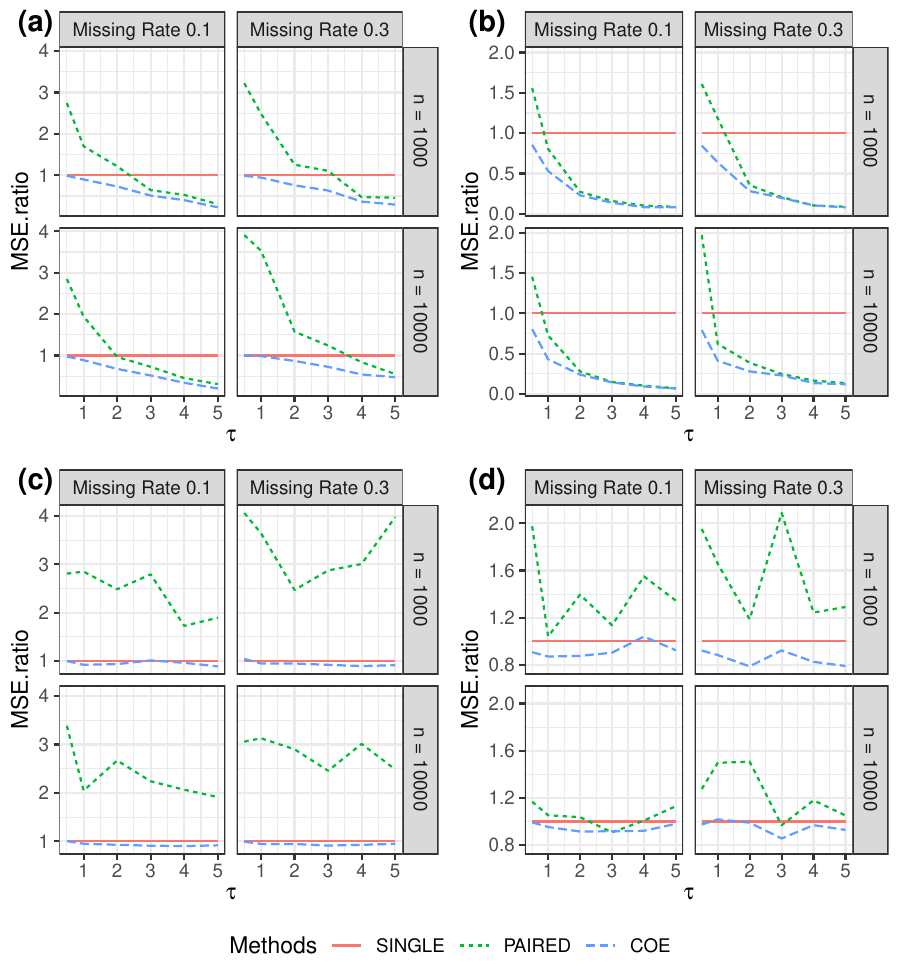}
\caption{Modified MSE.ratio in \eqref{eq:evaluationModified} under user effect settings (a)-(d) for integer/count responses.}\label{fig:count}
\end{figure}

Next we evaluate different methods using $\mathrm{MSE.ratio}$
in \eqref{eq:evaluationModified} with $\beta_1$ replaced by the actual estimand $\tilde\beta^l_1$, and both $\tilde{y}_{i1}(1)$ and $\tilde{y}_{i1}(-1)$ can be generated from the simulation model. 
The results for binary and integer/count outcomes are shown in Figures \ref{fig:binary} and \ref{fig:count}, respectively. 
The comparison conclusion is similar to the continuous outcome case in Figure \ref{fig:user}. 
The advantage of COE is robust over different types of responses. 
What is more, the advantage of COE over PAIRED is more significant than the continuous outcome case.

\section{Case Study}\label{sec:case}

\begin{figure}[t!]
\centering
        \includegraphics[scale=1]{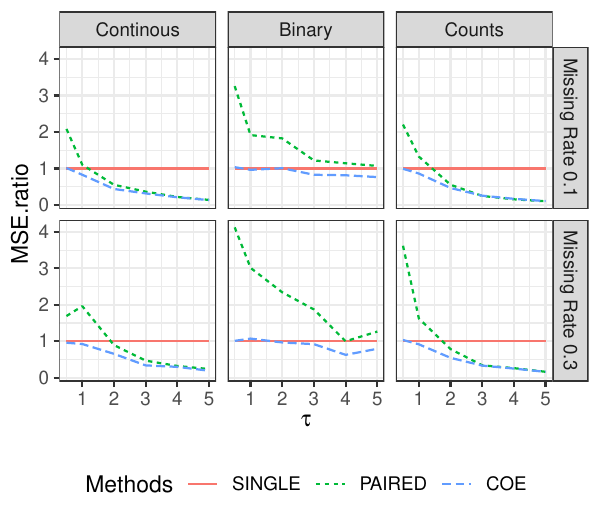}
\caption{MSE.ratio in \eqref{eq:evaluation} (for continuous responses) 
or modified MSE.ratio in \eqref{eq:evaluationModified} (for binary and count responses) with user effect extracted from real data.}\label{fig:real}
\end{figure}

We create a case study based on the customer personality dataset in Kaggle \citep{kaggledata}. %This dataset contains 2216 users with complete records \textcolor{red}{record on what?}. 
This dataset contains users' information and their responses to five campaigns. 
We take 2216 users with complete records of all the features. 
Through variable selection, we keep income, number of kids at home, and number of teens at home as effective user features in model fitting.  
The response for each of the five campaigns is a binary outcome,  indicating accept (1) or not (0). Let $\mathrm{Accept}_{i,k}$ be the status of the $i$-th user to the $k$-th campaign, and $\mathrm{income}_{i}$,  $\mathrm{kids}_{i}$ and $\mathrm{teens}_{i}$ be the covariates of $i$-th user. 
We model the binary campaign records by a generalized mixed-effect model:
\[
\Phi^{-1}\left\{\mathbb P(\mathrm{Accept}_{i,k}=1)\right\}=\alpha_k+\gamma_1\times \mathrm{income}_{i}+
\gamma_2\times \mathrm{kids}_{i}+\gamma_3\times \mathrm{teens}_{i}+\tilde{u}_i,
\]
where $\alpha_k$'s and $\gamma_j$'s are the fixed effects and $\tilde{u}_i$'s are the random user effects with mean zero and variance $\tau^2$. By fitting this model, we extract the user effect $u_i$ be the conditional mean of $\gamma_1\times \mathrm{income}_{i}+
\gamma_2\times \mathrm{kids}_{i}+\gamma_3\times \mathrm{teens}_{i}+\tilde{u}_i$ given data.  We then use the extracted user effects to generate responses under \eqref{eq:simulation_model} and convert to binary and count responses by \eqref{eq:binary} and \eqref{eq:integer}, respectively. 
The results are shown in Figure \ref{fig:real}, which demonstrates a similar comparison as in Section \ref{sec:num}.

Notice that, the case study is a pseudo-study based on real data. The reason is that our analysis approach can not be used on this real data directly, unfortunately. 
The five campaigns in the data correspond to five experiments, but there are no treatments assigned to the five experiments. Therefore, treatment effect estimation does not apply to the data. We use this dataset to extract users' effects (a linear combination of user covariates and additional random effects) and add a synthetic additive treatment effect to generate new responses. The three types of responses can be explained under the scope of the data as accepting campaign requests or not (the original 0-1 response), number of purchases (integer or count responses), and amount spent (continuous responses). Therefore, potentially the proposed collaborative analysis approach can be used on consumer platforms to assist the decision-making of different versions of services and advertising campaigns.

\section{Conclusion and Discussion}\label{sec:con}

This paper proposed a collaborative analysis approach for a pair of A/B testing experiments carried out on the same set of users. 
The proposed approach can work well in practical situations where partially paired A/B testing experiments are commonly encountered.
%motivated by fully paired A/B testing under balanced and orthogonal designs but can be extended to
Theoretical results show that the proposed collaborative estimators are asymptotically the best linear unbiased estimators under certain assumptions. Also, numerical results demonstrate that the proposed collaborative estimator is robust under a set of misspecified models.
Overall, compared to the linear mixed effect model, the proposed collaborative analysis approach is computationally efficient and easy to implement over online experimental platforms. Compared to single experiment analysis  and paired analysis approaches, the proposed collaborative analysis approach consistently gives more accurate estimators 
over different scenarios. 

Although this work focuses on paired A/B testing experiments, the proposed concept of collaborative analysis 
can be extended to the collaborative analysis of multiple A/B testing experiments that involve the same set of users. 
The idea of contrast can be used for such a generalization of the proposed method. 
It will be also interesting to consider multivariate mixed responses in each of the multiple A/B testing experiments. 
One can leverage some techniques developed in \cite{kang2018bayesian} and \cite{chen2023bayesian} to facilitate the model estimation and inference. 
Also, the experimental design issue in the presence of user covariates (e.g., \cite{li2021covariate})  and network information (e.g., \cite{pokhilko2019d, zhang2022locally} can be interesting to explore in the future. 
One can further extend the framework to personalized preference learning via collaborative experiments (e.g., \cite{zhang2022min,fisher2022bayesian,li2023optimal}).

\appendix
\section{Appendix}

\subsection{Proof of Proportion 1}\label{app:prop1}

Let $\bm y_1=(y_{1,1}, \ldots, y_{n,1})^\top$
and $\bm y_2=(y_{1,2}, \ldots, y_{n,2})^\top$
be the response vectors collected from the first and second experiments. 
According to the true model \eqref{eq:true_model},
the covariance matrix of the combined response vector $\bm y=(\bm y^\top_1, \bm y^\top_2)^\top$ is given by
\[
   \bm V = \left[
\begin{array}{rr}
(\sigma_1^2+\tau^2)\bm I_{n} & \tau^2 \bm I_{n}  \\
\tau^2 \bm I_{n} & (\sigma_2^2+\tau^2)\bm I_{n} 
\end{array}
\right]
\]
with
\[
\bm V^{-1} = \frac{1}{(\sigma_1^2+\tau^2)(\sigma_2^2+\tau^2)-\tau^4}\left[
\begin{array}{rr}
(\sigma_2^2+\tau^2)\bm I_{n} & -\tau^2 \bm I_{n}  \\
-\tau^2 \bm I_{n} & (\sigma_1^2+\tau^2)\bm I_{n} 
\end{array}
\right]
\]
and the covariates matrix is 
\[
\bm X=\left(\begin{array}{cccc}
{\bf 1}_{n} & {\bf 0} & \bm x_{1} & {\bf 0} \\
{\bf 0} & {\bf 1}_{n} & {\bf 0} & \bm x_{2} \\
\end{array}\right)
\]
with $\bm x_k=(x_{1,k}, \ldots, x_{n,k})^\top$ be the design of the $k$-th experiment for $k=1, 2$. 
Therefore, the weighted least squared estimator of the linear coefficients $\bm\theta=(\alpha_1, \alpha_2, \beta_1, \beta_2)^\top$ is
\begin{equation}\label{eq:lse_pair}
    \hat{\bm\theta}^{wls}=(\hat\alpha^{wls}_1, \hat\alpha^{wls}_2,
    \hat\beta^{wls}_1,
    \hat\beta^{wls}_2
    )^\top=(\bm X^\top \bm V^{-1}\bm X)^{-1}\bm X^\top \bm V^{-1}\bm y,
\end{equation}
with variance-covariance matrix
\begin{equation}\label{eq:var_lse_diff}
\mathrm{Var}(\hat{\bm\theta}^{wls})=(\bm X^\top \bm V^{-1}\bm X)^{-1}.
\end{equation}
Under the orthogonal assumption
\[
\bm X^\top \bm V^{-1}\bm X=\frac{1}{(\sigma_1^2+\tau^2)(\sigma_2^2+\tau^2)-\tau^4}\left(\begin{array}{cccc}
     n(\sigma^2_2+\tau^2)& -n\tau^2& 0 &0\\
  -n\tau^2 & n(\sigma^2_1+\tau^2)& 0& 0\\
   0& 0 & n(\sigma^2_2+\tau^2) &0\\
    0& 0 & 0&n(\sigma^2_1+\tau^2) \\
\end{array}\right).
\]
Also, 
\[
\bm X^\top \bm V^{-1}\bm y=\frac{1}{(\sigma_1^2+\tau^2)(\sigma_2^2+\tau^2)-\tau^4}\left(\begin{array}{c}
     (\sigma^2_2+\tau^2)\sum^n_{i=1}y_{i,1}-\tau^2\sum^n_{i=1}y_{i,2}\\
     (\sigma^2_1+\tau^2)\sum^n_{i=1}y_{i,2}-\tau^2\sum^n_{i=1}y_{i,1}\\
      (\sigma^2_2+\tau^2)\sum^n_{i=1}x_{i,1}y_{i,1}-\tau^2\sum^n_{i=1}x_{i,1}y_{i,2}\\
     (\sigma^2_1+\tau^2)\sum^n_{i=1}x_{i,2}y_{i,2}-\tau^2\sum^n_{i=1}x_{i,2}y_{i,1}
\end{array}\right)
\]
Therefore, 
\[
\hat\beta^{wls}_1=\frac{(\sigma^2_2+\tau^2)\sum^n_{i=1}x_{i,1}y_{i,1}-\tau^2\sum^n_{i=1}x_{i,1}y_{i,2}}{n(\sigma^2_2+\tau^2)}=\frac{\tau^2\hat\beta^p_1+\sigma^2_2\hat\beta^s_1}{\sigma^2_2+\tau^2}=\hat\beta^c_1,
\]
and
\[
\hat\beta^{wls}_2=\frac{(\sigma^2_1+\tau^2)\sum^n_{i=1}x_{i,1}y_{i,2}-\tau^2\sum^n_{i=1}x_{i,1}y_{i,1}}{n(\sigma^2_1+\tau^2)}=\frac{\tau^2\hat\beta^p_2+\sigma^2_1\hat\beta^s_2}{\sigma^2_1+\tau^2}=\hat\beta^c_2.
\]
Therefore, the hybrid estimators are
equivalent to the weighted least squared estimators, i.e., the best linear unbiased estimator under the given assumptions. Thus, the conclusion in part (i) holds.

Notice that 
\begin{align*}
\hat\beta^s_1 &=\beta_1+\frac{1}{n}\sum^n_{i=1}x_{i,1}(\varepsilon_{i,1}+u_i), \quad
\hat\beta^p_1 =\beta_1+\frac{1}{n}\sum^n_{i=1}x_{i,1}(\varepsilon_{i,1}-\varepsilon_{i,2})\\
\hat\beta^s_2 &=\beta_2+\frac{1}{n}\sum^n_{i=1}x_{i,2}(\varepsilon_{i,2}+u_i),\quad
\hat\beta^p_2 =\beta_2+\frac{1}{n}\sum^n_{i=1}x_{i,2}(\varepsilon_{i,2}-\varepsilon_{i,1}).
\end{align*}
Consider $(x_{i,1}(\varepsilon_{i,1}+u_i), x_{i,1}(\varepsilon_{i,1}-\varepsilon_{i,2}), x_{i,2}(\varepsilon_{i,2}+u_i), x_{i,2}(\varepsilon_{i,2}-\varepsilon_{i,1}))^\top$'s as independent random vectors, the Lindbergh-Feller multivariate central limit theorem gives that, as $n\rightarrow\infty$
\[
\frac{1}{\sqrt{n}}\left(\hat\beta^s_1-\beta_1, 
 \hat\beta^p_1-\beta_1, \hat\beta^s_2-\beta_2, \hat\beta^p_2-\beta_2 \right)^\top\rightarrow \mathcal N_4(\bm 0_4, \Sigma_4),
\]
in distribution with 
\[
\Sigma_4=\left[
\begin{array}{cccc}
\sigma^2_1+\tau^2 & \sigma^2_1 &0  & 0\\
\sigma^2_1 & \sigma^2_1+\sigma^2_2 & 0 & 0\\
0 & 0 & \sigma^2_2+\tau^2 & \sigma^2_2\\
0 & 0 & \sigma^2_2 & \sigma^2_1+\sigma^2_2\\
\end{array}
\right].
\]
We can directly find the joint asymptotic normal distribution of $\hat\beta^c_1-\beta_1$ and $\hat\beta^c_2-\beta_2$ as the linear combination of $\left(\hat\beta^s_1-\beta_1, 
 \hat\beta^p_1-\beta_1, \hat\beta^s_2-\beta_2, \hat\beta^p_2-\beta_2 \right)^\top$. This completes the proof of part (ii).

\subsection{Proof of Proportion 2}\label{app:prop2}
%Assume that 
%\[
%\frac{n_0}{n}\rightarrow r_0,\quad
%\frac{n_1}{n}\rightarrow r_1,\quad
%\frac{n_2}{n}\rightarrow r_2,
%\]
%where $r_0$, $r_1$ and $r_2$ are constants from $(0, 1)$. 
We first provide the weighted least squared estimators of $\beta_1$ and $\beta_2$.
Define the following notations: 
\begin{itemize}
\item $\bm y_{1,s}$: first experiment's output for the shared users with the 2nd experiment. 
\item $\bm y_{1,ns}$: first experiment's output for the non-shared users. 
\item $\bm y_{2,s}$: 2nd experiment's output for the shared users with the first experiment. 
\item $\bm y_{2,ns}$: 2nd expeirment's output for the non-shared users. 
\end{itemize}
Given Table \ref{tb1}, the sizes of the four vectors are $n_0$, $n_1$, $n_0$, and $n_2$, respectively. Similarly, we define the random effects $\bm u_s$ as the ones for the shared users, of size $n_0$, and $\bm u_{1,ns}$ and $\bm u_{2,ns}$ are the random effects of the non-shared groups for the two experiments, of size $n_1$ and $n_2$, respectively. 
There are also four random noise, $\bm \epsilon_{1,s}$, $\bm\epsilon_{1, ns}$, $\bm \epsilon_{2,s}$, and $\bm \epsilon_{2,ns}$. 
Based on the model assumptions in \eqref{eq:true_model}, we have that
\begin{equation*}
\left[\begin{array}{c}
\bm y_{1,s}\\
\bm y_{1,ns}\\
\bm y_{2,s}\\
\bm y_{2,ns}
\end{array}\right]=
\left[
\begin{array}{cccc}
{\bf 1}_{n_0} & {\bf 0} & \bm x_{1,s} & {\bf 0} \\
{\bf 1}_{n_1} & {\bf 0} & \bm x_{1,ns} & \\
{\bf 0} & {\bf 1}_{n_0} & {\bf 0} & \bm x_{2,s} \\
{\bf 0} & {\bf 1}_{n_2} & {\bf 0} & \bm x_{2,ns} \\
\end{array}
\right]
\left[\begin{array}{c}
\alpha_1\\
\alpha_2\\
\beta_1\\
\beta_2
\end{array}
\right]+
\left[\begin{array}{c}
\bm u_{s}\\
\bm u_{1,ns}\\
\bm u_{s}\\
\bm u_{2,ns}
\end{array}\right]+
\left[\begin{array}{c}
\bm \epsilon_{1,s}\\
\bm \epsilon_{1,ns}\\
\bm \epsilon_{2,s}\\
\bm \epsilon_{2,ns}
\end{array}\right]
\end{equation*}
The covariance matrix of the responses is
\[
\bm V= \left[
\begin{array}{rrrr}
(\sigma_1^2+\tau^2)\bm I_{n_0} & {\bf 0} & \tau^2 \bm I_{n_0} & {\bf 0} \\
{\bf 0} & (\sigma_1^2+\tau^2)\bm I_{n_1} & {\bf 0} & {\bf 0} \\
\tau^2 \bm I_{n_0} & {\bf 0} & (\sigma_2^2+\tau^2)\bm I_{n_0} & {\bf 0} \\
{\bf 0} & {\bf 0} & {\bf 0} & (\sigma^2+\tau^2)\bm I_{n_2}
\end{array}
\right]
\]
with inverse  $\bm V^{-1}$
{
\scriptsize
\[\left[
\begin{array}{rrrr}
\left(\sigma_1^2+\frac{\sigma_2^2\tau^2}{\sigma_2^2+\tau^2}\right)^{-1}\bm I_{n_0} & {\bf 0} & -\left[(\sigma_1^2+\tau^2)(\sigma_2^2+\tau^2)/\tau^2-\tau^2\right]^{-1} \bm I_{n_0} & {\bf 0} \\
{\bf 0} & (\sigma_1^2+\tau^2)^{-1}\bm I_{n_1} & {\bf 0} & {\bf 0} \\
 -\left[(\sigma_1^2+\tau^2)(\sigma_2^2+\tau^2)/\tau^2-\tau^2\right]^{-1} \bm I_{n_0} & {\bf 0} & \left[(\sigma_2^2+\tau^2)-\frac{\tau^4}{\sigma_1^2+\tau^2}\right]^{-1}\bm I_{n_0} & {\bf 0} \\
{\bf 0} & {\bf 0} & {\bf 0} & (\sigma_2^2+\tau^2)^{-1}\bm I_{n_2}
\end{array}
\right]
\]}
Then we have that $\bm X^\top \bm V^{-1} \bm X=$
\[
%\bm X^\top \bm V^{-1} \bm X=
\left[
\begin{array}{rrrr}
(a n_0+b n_1) & e n_0 & a {\bf 1}_{n_0}^\top \bm x_{1,s}+b {\bf 1}_{n_1}^\top\bm x_{1,ns} & e {\bf 1}_{n_0}^\top \bm x_{2,s}\\
e n_0 & c n_0+d n_2 & e {\bf 1}_{n_0}^\top\bm x_{1,s} & c{\bf 1}_{n_0}^\top \bm x_{2,s}+ d c{\bf 1}_{n_2}^\top \bm x_{2,ns}\\
a {\bf 1}_{n_0}^\top \bm x_{1,s}+b {\bf 1}_{n_1}^\top\bm x_{1,ns} & e {\bf 1}_{n_0}^\top\bm x_{1,s} & (a n_0+b n_1) & e \bm x_{1,s}^\top \bm x_{2,s} \\
e {\bf 1}_{n_2}^\top \bm x_{2,s} & c{\bf 1}_{n_0}^\top \bm x_{2,s}+ d c{\bf 1}_{n_2}^\top \bm x_{2,ns} & e \bm x_{1,s}^\top \bm x_{2,s} & c n_0+d n_2
\end{array}
\right].
\]
Here $a=\left(\sigma_1^2+\frac{\sigma_2^2\tau^2}{\sigma_2^2+\tau^2}\right)^{-1}$, $b=(\sigma_1^2+\tau^2)^{-1}$, $c=\left[(\sigma_2^2+\tau^2)-\frac{\tau^4}{\sigma_1^2+\tau^2}\right]^{-1}$, $d=(\sigma_2^2+\tau^2)^{-1}$, and $e=-\left[(\sigma_1^2+\tau^2)(\sigma_2^2+\tau^2)/\tau^2-\tau^2\right]^{-1}$. Also, we have that
\[
\bm X^\top \bm V^{-1} \bm y=\left[
\begin{array}{l}
a {\bf 1}_{n_0}^\top \bm y_{1,s}+b {\bf 1}_{n_1}^\top \bm y_{1,ns}+e {\bf 1}_{n_0}^\top \bm y_{2,s}\\
e {\bf 1}_{n_0}^\top \bm y_{1,s}+c{\bf 1}_{n_0}^\top \bm y_{2,s}+d {\bf 1}_{n_2}^\top \bm y_{2,ns}\\
a \bm x_{1,s}^\top \bm y_{1,s}+b \bm x_{1,ns}^\top \bm y_{1,ns}+e \bm x_{1,s}^\top \bm y_{2,s}\\
e \bm x_{2,s}^\top \bm y_{1,s}+c\bm x_{2,s}^\top \bm y_{2,s}+d \bm x_{2,ns}^\top \bm y_{2,ns}
\end{array}
\right].
\]
Then the weighted least squared estimator is given by
\[
(\hat\alpha^{wls}_1, \hat\alpha^{wls}_2,
    \hat\beta^{wls}_1,
    \hat\beta^{wls}_2
    )^\top=\left(\bm X^\top \bm V^{-1} \bm X\right)^{-1}\bm X^\top \bm V^{-1} \bm y.
\]
Let
\begin{align*}
\bm A&=\left[\begin{array}{rr}
a n_0+b n_1 & e n_0\\ 
e n_0 & c n_0+d n_2 
\end{array}
\right]\\
\bm B&=\left[\begin{array}{rr}
 a {\bf 1}_{n_0}^\top \bm x_{1,s}+b {\bf 1}_{n_1}^\top\bm x_{1,ns} & e {\bf 1}_{n_0}^\top \bm x_{2,s}\\
 e {\bf 1}_{n_0}^\top\bm x_{1,s} & c{\bf 1}_{n_0}^\top \bm x_{2,s}+ d c{\bf 1}_{n_2}^\top \bm x_{2,ns}
\end{array}
\right]\\
\bm D&=\left[\begin{array}{rr}
 a n_0+b n_1 & e \bm x_{1,s}^\top \bm x_{2,s} \\
e \bm x_{1,s}^\top \bm x_{2,s} & c n_0+d n_2
\end{array}
\right].
\end{align*}
We have that
\[
\mathrm{Var}\left(\left[\begin{array}{r}\hat\beta^{wle}_1\\\hat\beta^{wle}_2\end{array}\right]\right)=\left(\bm D-\bm B^\top \bm A^{-1}\bm B\right)^{-1}
\]
Thus
\[
\left[n\mathrm{Var}\left(\left[\begin{array}{r}\hat\beta^{wle}_1\\\hat\beta^{wle}_2\end{array}\right]\right)\right]^{-1}=\frac{1}{n}\bm D-\left(\frac{1}{n}\bm B^\top\right) \left(\frac{1}{n}\bm A\right)^{-1}\left(\frac{1}{n}\bm B\right)
\]
Note that, under Assumption \ref{ass:n}, as $n\rightarrow\infty$
\[
\frac{1}{n}\bm A\rightarrow \left[\begin{array}{rr}
a r_0+b r_1 & e r_0\\ 
e r_0 & c r_0+d r_2 
\end{array}
\right]
\]
and
\[
\frac{1}{n}\bm D\rightarrow \left[\begin{array}{rr}
 a r_0+b r_1 & 0 \\
0 & c r_0+d r_2
\end{array}
\right]
\]
which are constant matrices. Under 
Assumption \ref{ass:nearly}, we have that 
\[
\frac{1}{n}\bm B\rightarrow \bm 0
\]
as $n\rightarrow\infty$.
Therefore, 
\[
\left[n\mathrm{Var}\left(\left[\begin{array}{r}\hat\beta^{wle}_1\\\hat\beta^{wle}_2\end{array}\right]\right)\right]^{-1}\rightarrow
\left[\begin{array}{rr}
 a r_0+b r_1 & 0 \\
0 & c r_0+d r_2
\end{array}
\right]
\]
and
\begin{equation}\label{eq:avar}
n\mathrm{Var}\left(\left[\begin{array}{r}\hat\beta^{wle}_1\\\hat\beta^{wle}_2\end{array}\right]\right)\rightarrow
\left[\begin{array}{rr}
 1/(a r_0+b r_1) & 0 \\
0 & 1/(c r_0+d r_2)
\end{array}
\right]
\end{equation}
as $n\rightarrow\infty$. 
Also, we have that
\begin{align}
\left[\begin{array}{r}\hat\beta^{wle}_1\\\hat\beta^{wle}_2\end{array}\right]&=\mathrm{Var}\left(\left[\begin{array}{r}\hat\beta^{wle}_1\\\hat\beta^{wle}_2\end{array}\right]\right)\left[\begin{array}{rr}-\bm B^\top \bm A^{-1} & \bm I_2\end{array}\right]\bm X^\top \bm V^{-1} \bm y\\
&= n\mathrm{Var}\left(\left[\begin{array}{r}\hat\beta^{wle}_1\\\hat\beta^{wle}_2\end{array}\right]\right)\left[\begin{array}{rr}-\left(n^{-1}\bm B^\top\right) \left(n^{-1}\bm A\right)^{-1} & \bm I_2\end{array}\right]\frac{1}{n}\bm X^\top \bm V^{-1} \bm y
\end{align}

Also, we can express the collaborative estimators by
\[
\hat{\beta}^c_1=\frac{-e \bm x_{1,s}^\top (\bm y_{1,s}-\bm y_{2,s})+b \bm x_{1,ns}^\top \bm y_{1,ns}+(a+e) \bm x_{1,s}^\top \bm y_{1,s}}{an_0+bn_1}
=\frac{a \bm x_{1,s}^\top \bm y_{1,s}+b \bm x_{1,ns}^\top \bm y_{1,ns}+e \bm x_{1,s}^\top \bm y_{2,s}}{an_0+bn_1},
\]
and
\[
\hat{\beta}^c_2=\frac{-e \bm x_{2,s}^\top (\bm y_{2,s}-\bm y_{1,s})+d \bm x_{2,ns}^\top \bm y_{2,ns}+(c+e) \bm x_{2,s}^\top \bm y_{2,s}}{cn_0+dn_2}
=\frac{e \bm x_{2,s}^\top \bm y_{1,s}+c\bm x_{2,s}^\top \bm y_{2,s}+d \bm x_{2,ns}^\top \bm y_{2,ns}}{cn_0+dn_2}.
\]
We can then express
\[
\left[\begin{array}{r}\hat\beta^{c}_1\\\hat\beta^{c}_2\end{array}\right]=n\left[\begin{array}{rrrr}
0 & 0& 1/(a n_0+b n_1) & 0 \\
0 & 0& 0 & 1/(c n_0+d n_2)
\end{array}
\right]\frac{1}{n}\bm X^\top \bm V^{-1} \bm y
\]
Let 
\[
\bm M_n=
n\mathrm{Var}\left(\left[\begin{array}{r}\hat\beta^{wle}_1\\\hat\beta^{wle}_2\end{array}\right]\right)\left[\begin{array}{rr}
\left(n^{-1}\bm B^\top\right) \left(n^{-1}\bm A\right)^{-1} & \bm I_2\end{array}\right]\\
-n\left[\begin{array}{rrrr}
0 & 0& 1/(a n_0+b n_1) & 0 \\
0 & 0& 0 & 1/(c n_0+d n_2)
\end{array}
\right]
\]
Then $\bm M_n\rightarrow 0$ as $n\rightarrow\infty$.
Therefore, 
\[
\left[\begin{array}{r}\hat\beta^{wle}_1\\\hat\beta^{wle}_2\end{array}\right]-\left[\begin{array}{r}\hat\beta^{c}_1\\\hat\beta^{c}_2\end{array}\right]=\bm M_n\cdot\frac{1}{n}\bm X^\top \bm V^{-1} \bm y\rightarrow 0
\]
as $n\rightarrow\infty$, which shows the asymptotic equivalence between the weighted least squared estimators and the collaborative estimators. 
This completes the proof of part (i).

For part (ii), let $r_{i,k}$ be a 0-1 value indicating whether or not 
the $k$-th response from the $i$-th user is available. Therefore, we can express 
\begin{align*}
n^{-1}_0\sum^{n_0}_{i=1}x_{i,1}y_{i,1} &=\frac{n}{n_0}\cdot n^{-1}\sum^{n}_{i=1}x_{i,1}r_{i,1}r_{i,2}y_{i,1}=\beta_1+
\frac{n}{n_0}\cdot \frac{\sum^{n_0}_{i=1}x_{i,1}}{n}\alpha+\frac{n}{n_0}\cdot n^{-1}\sum^{n}_{i=1}x_{i,1}r_{i,1}r_{i,2}(\varepsilon_{i,1}+u_i)\\
n^{-1}_0\sum^{n_0}_{i=1}x_{i,1}z_{i} &=\beta_1+
\frac{n}{n_0}\cdot \frac{\sum^{n_0}_{i=1}x_{i,1}}{n}\alpha-\frac{n}{n_0}\cdot \frac{\sum^{n_0}_{i=1}x_{i,1}x_{i,2}}{n}\beta_2+\frac{n}{n_0}\cdot n^{-1}\sum^{n}_{i=1}x_{i,1}r_{i,1}r_{i,2}(\varepsilon_{i,1}-\varepsilon_{i,2})\\
n^{-1}_1\sum^{n_0+n_1}_{i=n_0+1}x_{i,1}y_{i,1} &=
%\frac{n}{n_1}\cdot n^{-1}\sum^{n}_{i=1}x_{i1}r_{i1}(1-r_{i2})y_{i1}=
\beta_1+
\frac{n}{n_1}\cdot \frac{\sum^{n_0+n_1}_{i=n_0+1}x_{i,1}}{n}\alpha+\frac{n}{n_1}\cdot n^{-1}\sum^{n}_{i=1}x_{i,1}r_{i,1}(1-r_{i,2})(\varepsilon_{i,1}+u_i)
\end{align*}
also, for estimating $\beta_2$, we have
\begin{align*}
n^{-1}_0\sum^{n_0}_{i=1}x_{i,2}y_{i,2} &=\frac{n}{n_0}\cdot n^{-1}\sum^{n}_{i=1}x_{i,2}r_{i,1}r_{i,2}y_{i,2}=\beta_2+
\frac{n}{n_0}\cdot \frac{\sum^{n_0}_{i=1}x_{i,2}}{n}\alpha+\frac{n}{n_0}\cdot n^{-1}\sum^{n}_{i=1}x_{i,2}r_{i,1}r_{i,2}(\varepsilon_{i,2}+u_i)\\
n^{-1}_0\sum^{n_0}_{i=1}x_{i,2}z_{i} &=\beta_2+
\frac{n}{n_0}\cdot \frac{\sum^{n_0}_{i=1}x_{i,2}}{n}\alpha-\frac{n}{n_0}\cdot \frac{\sum^{n_0}_{i=1}x_{i,1}x_{i,2}}{n}\beta_1+\frac{n}{n_0}\cdot n^{-1}\sum^{n}_{i=1}x_{i,2}r_{i,1}r_{i,2}(\varepsilon_{i,2}-\varepsilon_{i,1})\\
n^{-1}_2\sum^{n_0+n_1+n_2}_{i=n_0+n_1+1}x_{i,2}y_{i,2} &=
%\frac{n}{n_1}\cdot n^{-1}\sum^{n}_{i=1}x_{i1}r_{i1}(1-r_{i2})y_{i1}=
\beta_2+
\frac{n}{n_2}\cdot \frac{\sum^{n_0+n_1+n_2}_{i=n_0+n_1+1}x_{i,2}}{n}\alpha+\frac{n}{n_2}\cdot n^{-1}\sum^{n}_{i=1}x_{i,2}r_{i,2}(1-r_{i,1})(\varepsilon_{i,2}+u_i)
\end{align*}
First, we apply the Lindbergh-Feller's multivariate central limit theorem to $n^{-1}\sum^{n}_{i=1}x_{i,1}r_{i,1}r_{i,2}(\varepsilon_{i,1}+u_i)$, $n^{-1}\sum^{n}_{i=1}x_{i,1}r_{i,1}r_{i,2}(\varepsilon_{i,1}-\varepsilon_{i,2})$, $n^{-1}\sum^{n}_{i=1}x_{i,1}r_{i,1}(1-r_{i,2})(\varepsilon_{i,1}+u_i)$, $n^{-1}\sum^{n}_{i=1}x_{i,2}r_{i,1}r_{i,2}(\varepsilon_{i,2}+u_i)$, $n^{-1}\sum^{n}_{i=1}x_{i,2}r_{i,1}r_{i,2}(\varepsilon_{i,2}-\varepsilon_{i,1})$, and $n^{-1}\sum^{n}_{i=1}x_{i,2}r_{i,2}(1-r_{i,1})(\varepsilon_{i,2}+u_i)$. Under Assumptions \ref{ass:n}-\ref{ass:nearly}, the resulting multivariate asymptotic distribution has mean zero and variance-covariance matrix:
\[
\left[\begin{array}{cccccc}
r_0 (\sigma^2_1+\tau^2) & r_0 \sigma^2_1& 0 & 0 & 0 &0\\
r_0 \sigma^2_1 & r_0(\sigma^2_1+\sigma^2_2) & 0 & 0 & 0 &0\\
0&0&r_1(\sigma^2_1+\tau^2)& 0 & 0 &0\\
0&0&0  &r_0(\sigma^2_2+\tau^2) & r_0\sigma^2_2 &0\\
0&0&0  & r_0\sigma^2_2) & r_0(\sigma^2_1+\sigma^2_2) &0\\
0&0&0& 0 & 0 &r_2(\sigma^2_2+\tau^2)\\
\end{array}\right].
\]
Then we have that
\[
\sqrt{n}\left[\begin{array}{c}
\frac{\sum^{n_0}_{i=1}x_{i,1}y_{i,1}}{n_0}-\beta_1\\
\frac{\sum^{n_0}_{i=1}x_{i,1}z_{i}}{n_0}-\beta_1\\
\frac{\sum^{n_0+n_1}_{i=n_0+1}x_{i,1}y_{i,1}}{n_1}-\beta_1\\
\frac{\sum^{n_0}_{i=1}x_{i,2}y_{i,2}}{n_0}-\beta_2\\
\frac{\sum^{n_0}_{i=1}x_{i,2}z_{i}}{n_0}-\beta_2\\
\frac{\sum^{n_0+n_1+n_2}_{i=n_0+n+1+1}x_{i,2}y_{i,2}}{n_1}-\beta_2\\
\end{array}\right]\xrightarrow{d} \mathcal N_6\left(\bm 0_6, \bm V_6\right),
\]
with 
\[
\left[\begin{array}{cccccc}
r^{-1}_0 (\sigma^2_1+\tau^2) & r^{-1}_0 \sigma^2_1& 0 & 0 & 0 &0\\
r^{-1}_0 \sigma^2_1 & r^{-1}_0(\sigma^2_1+\sigma^2_2) & 0 & 0 & 0 &0\\
0&0&r^{-1}_1(\sigma^2_1+\tau^2)& 0 & 0 &0\\
0&0&0  &r^{-1}_0(\sigma^2_2+\tau^2) & r^{-1}_0\sigma^2_2 &0\\
0&0&0  & r^{-1}_0\sigma^2_2 & r^{-1}_0(\sigma^2_1+\sigma^2_2) &0\\
0&0&0& 0 & 0 &r^{-1}_2(\sigma^2_2+\tau^2)\\
\end{array}\right],
\]
which indicates the joint asymptotic normal distribution of $\hat\beta^c_1$ and $\hat\beta^c_2$.

\bibliographystyle{asa}
\bibliography{Ref}

\begin{thebibliography}{30}
\newcommand{\enquote}[1]{``#1''}
\expandafter\ifx\csname natexlab\endcsname\relax\def\natexlab#1{#1}\fi

\bibitem[{Andrzej and Tomasz(2012)}]{andrzej2012linear}
Andrzej, G. and Tomasz, B. (2012), \enquote{Linear Mixed Effects Models Using
  R: A Step-by-step Approach,} .

\bibitem[{Bates et~al.(2015)Bates, M{\"a}chler, Bolker, and Walker}]{lme4}
Bates, D., M{\"a}chler, M., Bolker, B., and Walker, S. (2015), \enquote{Fitting
  Linear Mixed-Effects Models Using {lme4},} \textit{Journal of Statistical
  Software}, 67, 1--48.

\bibitem[{Chen et~al.(2015)Chen, Ding, Geng, and Zhou}]{chen2015semiparametric}
Chen, H., Ding, P., Geng, Z., and Zhou, X.-H. (2015), \enquote{Semiparametric
  Inference of the Complier Average Causal Effect with Nonignorable Missing
  Outcomes,} \textit{ACM Transactions on Intelligent Systems and Technology
  (TIST)}, 7, 1--15.

\bibitem[{Chen et~al.(2023)Chen, Kang, Jin, and Deng}]{chen2023bayesian}
Chen, X., Kang, X., Jin, R., and Deng, X. (2023), \enquote{Bayesian Sparse
  regression for mixed multi-responses with application to runtime metrics
  prediction in fog manufacturing,} \textit{Technometrics}, 65, 206--219.

\bibitem[{Deng et~al.(2013)Deng, Xu, Kohavi, and Walker}]{deng2013improving}
Deng, A., Xu, Y., Kohavi, R., and Walker, T. (2013), \enquote{Improving the
  sensitivity of online controlled experiments by utilizing pre-experiment
  data,} in \textit{Proceedings of the sixth ACM international conference on
  Web search and data mining}, pp. 123--132.

\bibitem[{Ding(2024)}]{ding2024first}
Ding, P. (2024), \textit{A first course in causal inference}, CRC Press.

\bibitem[{Fisher et~al.(2022)Fisher, Zhang, Song, Gorsich, Hartman, and
  Skowronska}]{fisher2022bayesian}
Fisher, W., Zhang, Q., Song, Y., Gorsich, D., Hartman, G., and Skowronska, A.
  (2022), \enquote{Bayesian Sequential Preference Elicitation: A Tradespace
  Exploration Framework with Applications in Vehicle Concept Design,} in
  \textit{IIE Annual Conference. Proceedings}, Institute of Industrial and
  Systems Engineers (IISE), pp. 1--6.

\bibitem[{Gupta et~al.(2019)Gupta, Kohavi, Tang, Xu, Andersen, Bakshy, Cardin,
  Chandran, Chen, Coey, et~al.}]{gupta2019top}
Gupta, S., Kohavi, R., Tang, D., Xu, Y., Andersen, R., Bakshy, E., Cardin, N.,
  Chandran, S., Chen, N., Coey, D., et~al. (2019), \enquote{Top challenges from
  the first practical online controlled experiments summit,} \textit{ACM SIGKDD
  Explorations Newsletter}, 21, 20--35.

\bibitem[{Heffernan and Heffernan(2014)}]{heffernan2014assistments}
Heffernan, N.~T. and Heffernan, C.~L. (2014), \enquote{The ASSISTments
  ecosystem: Building a platform that brings scientists and teachers together
  for minimally invasive research on human learning and teaching,}
  \textit{International Journal of Artificial Intelligence in Education}, 24,
  470--497.

\bibitem[{Imbens and Rubin(2015)}]{Imbens_Rubin_2015}
Imbens, G.~W. and Rubin, D.~B. (2015), \textit{Causal Inference for Statistics,
  Social, and Biomedical Sciences: An Introduction}, Cambridge University
  Press.

\bibitem[{Jin and Ba(2023)}]{jin2023toward}
Jin, Y. and Ba, S. (2023), \enquote{Toward optimal variance reduction in online
  controlled experiments,} \textit{Technometrics}, 65, 231--242.

\bibitem[{Kang et~al.(2018)Kang, Kang, Deng, and Jin}]{kang2018bayesian}
Kang, L., Kang, X., Deng, X., and Jin, R. (2018), \enquote{A Bayesian
  hierarchical model for quantitative and qualitative responses,}
  \textit{Journal of Quality Technology}, 50, 290--308.

\bibitem[{Kohavi et~al.(2020)Kohavi, Tang, and Xu}]{kohavi2020trustworthy}
Kohavi, R., Tang, D., and Xu, Y. (2020), \textit{Trustworthy online controlled
  experiments: A practical guide to a/b testing}, Cambridge University Press.

\bibitem[{Larsen et~al.(2023)Larsen, Stallrich, Sengupta, Deng, Kohavi, and
  Stevens}]{larsen2023statistical}
Larsen, N., Stallrich, J., Sengupta, S., Deng, A., Kohavi, R., and Stevens,
  N.~T. (2023), \enquote{Statistical Challenges in Online Controlled
  Experiments: A Review of A/B Testing Methodology,} \textit{The American
  Statistician}, 0, 1--15.

\bibitem[{Li et~al.(2021)Li, Kang, and Huang}]{li2021covariate}
Li, Y., Kang, L., and Huang, X. (2021), \enquote{Covariate balancing based on
  kernel density estimates for controlled experiments,} \textit{Statistical
  Theory and Related Fields}, 5, 102--113.

\bibitem[{Li et~al.(2023)Li, Zhang, Khademi, and Yang}]{li2023optimal}
Li, Y., Zhang, Q., Khademi, A., and Yang, B. (2023), \enquote{Optimal Design of
  Controlled Experiments for Personalized Decision Making in the Presence of
  Observational Covariates,} \textit{The New England Journal of Statistics in
  Data Science}, 1, 386--393.

\bibitem[{Nassi and Jewkes(2021)}]{nassi_jewkes_2021}
Nassi, T. and Jewkes, H. (2021), \enquote{Simultaneous Experimentation: Run
  Multiple A/B Tests Concurrently,}
  \url{https://www.split.io/blog/simultaneous-experiments/}.

\bibitem[{Neyman(1923)}]{neyman1923application}
Neyman, J. (1923), \enquote{On the application of probability theory to
  agricultural experiments. Essay on principles,} \textit{Ann. Agricultural
  Sciences}, 1--51.

\bibitem[{Patel(2021)}]{kaggledata}
Patel, A. (2021), \enquote{Customer Personality Analysis,}
  https://www.kaggle.com/datasets/ imakash3011/customer-personality-analysis.

\bibitem[{Pokhilko et~al.(2019)Pokhilko, Zhang, Kang, et~al.}]{pokhilko2019d}
Pokhilko, V., Zhang, Q., Kang, L., et~al. (2019), \enquote{D-optimal design for
  network A/B testing,} \textit{Journal of Statistical Theory and Practice},
  13, 1--23.

\bibitem[{Poyarkov et~al.(2016)Poyarkov, Drutsa, Khalyavin, Gusev, and
  Serdyukov}]{poyarkov2016boosted}
Poyarkov, A., Drutsa, A., Khalyavin, A., Gusev, G., and Serdyukov, P. (2016),
  \enquote{Boosted decision tree regression adjustment for variance reduction
  in online controlled experiments,} in \textit{Proceedings of the 22nd ACM
  SIGKDD International Conference on Knowledge Discovery and Data Mining}, pp.
  235--244.

\bibitem[{Rubin(1974)}]{rubin1974estimating}
Rubin, D.~B. (1974), \enquote{Estimating causal effects of treatments in
  randomized and nonrandomized studies,} \textit{Journal of educational
  Psychology}, 66, 688.

\bibitem[{Selent et~al.(2016)Selent, Patikorn, and
  Heffernan}]{selent2016assistments}
Selent, D., Patikorn, T., and Heffernan, N. (2016), \enquote{Assistments
  dataset from multiple randomized controlled experiments,} in
  \textit{Proceedings of the Third (2016) ACM Conference on Learning@ Scale},
  pp. 181--184.

\bibitem[{Shen et~al.(2023)Shen, Mao, Zhang, Chen, Nie, and
  Deng}]{shen_mao_zhang_chen_nie_deng_2023}
Shen, S., Mao, H., Zhang, Z., Chen, Z., Nie, K., and Deng, X. (2023),
  \enquote{Clustering-Based Imputation for Dropout Buyers in Large-Scale Online
  Experimentation,} \textit{The New England Journal of Statistics in Data
  Science}, 1, 415--425.

\bibitem[{Syrgkanis et~al.(2019)Syrgkanis, Lei, Oprescu, Hei, Battocchi, and
  Lewis}]{syrgkanis2019machine}
Syrgkanis, V., Lei, V., Oprescu, M., Hei, M., Battocchi, K., and Lewis, G.
  (2019), \enquote{Machine learning estimation of heterogeneous treatment
  effects with instruments,} \textit{Advances in Neural Information Processing
  Systems}, 32.

\bibitem[{Zhang and Kang(2022)}]{zhang2022locally}
Zhang, Q. and Kang, L. (2022), \enquote{Locally optimal design for a/b tests in
  the presence of covariates and network dependence,} \textit{Technometrics},
  64, 358--369.

\bibitem[{Zhang et~al.(2022)Zhang, Khademi, and Song}]{zhang2022min}
Zhang, Q., Khademi, A., and Song, Y. (2022), \enquote{Min-max optimal design of
  two-armed trials with side information,} \textit{INFORMS Journal on
  Computing}, 34, 165--182.

\bibitem[{Zhao and Ding(2022)}]{zhao2022adjust}
Zhao, A. and Ding, P. (2022), \enquote{To adjust or not to adjust? estimating
  the average treatment effect in randomized experiments with missing
  covariates,} \textit{Journal of the American Statistical Association}, 1--11.

\bibitem[{Zhao et~al.(2024)Zhao, Ding, and Li}]{zhao2024covariate}
Zhao, A., Ding, P., and Li, F. (2024), \enquote{Covariate adjustment in
  randomized experiments with missing outcomes and covariates,}
  \textit{Biometrika}, asae017.

\bibitem[{Zuo et~al.(2024)Zuo, Ghosh, Ding, and Yang}]{zuo2024mediation}
Zuo, S., Ghosh, D., Ding, P., and Yang, F. (2024), \enquote{Mediation analysis
  with the mediator and outcome missing not at random,} \textit{Journal of the
  American Statistical Association}, 1--21.

\end{thebibliography}
\end{document}